\documentclass[noeprint,notitlepage,nofootinbib,10pt,showkeys]{revtex4-1}
\usepackage{graphicx}
\usepackage{amsmath}
\usepackage{amssymb}
\usepackage{array}
\usepackage[colorlinks]{hyperref}
\newcommand{\ub}{\text{ }\mu\text{b}}

\newcommand{\pb}{\text{ pb}}
\newcommand{\fb}{\text{ fb}}

\newcommand{\gev}{\text{ GeV}}
\newcommand{\tev}{\text{ TeV}}
\newcommand{\chip}[1]{\tilde{\chi}^+_#1}

\newcommand{\sqrtHatS}{\sqrt{\hat s}}

\usepackage{lineno}
\begin{document}

\title{Pair production of the lightest chargino at $\gamma\gamma$-collider}

\author{Nasuf Sonmez}
\email{nasuf.sonmez@ege.edu.tr}
\affiliation{Department of Physics, Faculty of Science, Ege University, 35040 Izmir, Turkey}
\date{\today}

%%%%%%%%%%%%%%%%%%%%%%%%%%%%%%%%%%%%%%%%%%%%%%%%%%%%%%

\begin{abstract}
Chargino pair production via photon-photon collision is investigated in the Minimal Supersymmetric Standard Model at a future linear collider. The process is computed using all the possible diagrams at the next-to-leading order, including box, triangle, and self-energy diagrams. The numerical analysis is carried out for the production rates of the lightest chargino pair in RNS, NS, mSUGRA, BB, and NUGM scenarios. These distinct benchmark models were introduced in the light of the LHC results presented at $\sqrt{s}=7-8\text{ TeV}$. Among these scenarios, the RNS has the highest production rate for the  $\gamma\gamma\rightarrow \tilde\chi_1^+ \tilde\chi_1^-$. The partonic cross section reaches up to $3.94\text{ pb}$ at $\sqrt{\hat{s}}=310\text{ GeV}$. The total convoluted cross section with the photon luminosity in a $e^+e^-$ machine is calculated as a function of the center-of-mass energy up to $1\text{ TeV}$. The convoluted cross section is $\sim1.05\text{ pb}$ at $\sqrt{s}=350\text{ GeV}$ depending on the polarization of the initial electron and laser polarization. The RNS along with the BB scenarios are accessible at $\sqrt{s}=500\text{ GeV}$ with a $\gamma\gamma$ collision mode on $e^+e^-$-collider.
\end{abstract}

%\begin{keyword}
%Supersymmetry; Lightest Chargino pair; Photon-photon collider; Benchmark points; Real photon correction
%\PACS 11.30.Pb \sep 14.80.Ly \sep 14.80.Nb
%\end{keyword}

\maketitle

%%%%%%%%%%%%%%%%%%%%%%%%%%%%%%%%%%%%%%%%%%%%%%%%%%%%%%%%%%%%%%%%%%%%%%%%
%%%%%%%%%%%%%%%%%%%%%%%%%%%%%%%%%%%%%%%%%%%%%%%%%%%%%%%%%%%%%%%%%%%%%%%%

\section{Introduction}
    
    The discovery of the Higgs boson at the LHC \cite{Aad:2012tfa, Chatrchyan:2012xdj, Aad:2015zhl} is a strong confirmation of the electroweak symmetry breaking mechanism in the Standard Model (SM). The SM explains the interactions between what is called the fundamental particles at the weak scale. In the last couple of decades, a substantial amount of literature about a higher scale was produced. In general, they are called beyond the Standard Model (BSM), and they propose new physics at higher scales. Among all the proposals of BSM physics, supersymmetry (Susy) is the one that provides a description of the strong and the electroweak interactions from Planck scale down to the weak scale; moreover, the quadratic divergences arising from the self-interaction of the scalar fields are reduced to logarithmic ones. The Minimal Supersymmetric Standard Model (MSSM)\cite{Martin:1997ns} is the minimal extension that defines a correspondence between bosons and fermions. According to the model, if a symmetry called the R-parity is conserved in the decay of superparticles, the lightest supersymmetric particle (LSP) becomes a natural candidate for a weakly-interacting dark matter. Besides the LSP, charginos are another hypothetical particles which drew attention. They are solely the fermionic mass eigenstates of the supersymmetric partners of the $W^\pm$ boson and the charged Higgses $H_{1,2}^\pm$. Measuring their masses and production rates could provide the possibility of determining the gaugino and the higgsino couplings. In the light of the results presented by the LHC in pp collisions at $\sqrt{s}=7,8,13\tev$, strict limits on Susy parameter space and constraints are set on the masses of sparticles \cite{Khachatryan:2014mma, Khachatryan:2014qwa, Khachatryan:2014doa, Altunkaynak:2015kia, Aaboud:2018zjf}.
    
    The particle physics community is preparing itself for the next collider projects in the last couple of decades. It is certain that the next collider will be a lepton-lepton collider - particularly for studying the properties of the Higgs boson - and currently there are several proposals. These are the Circular Electron-Positron Collider (CEPC, $\sqrt{s}=240\gev$) in China \cite{Gao:2017ubc, Liang:2016mue, Xiao:2015vrz}, the Future Circular Collider (FCC-ee, $\sqrt{s}=350\gev$) \cite{fcc-ee} at CERN \cite{Gomez-Ceballos:2013zzn}, and the Linear Collider Collaboration (International Linear Collaboration and CLIC projects with $\sqrt{s}=500\gev$) probably in Japan \cite{Yamamoto:2017lnu}. In all these proposals, the electron-positron collisions are planned. However, a $e^+e^-$-collider could also operate as a $\gamma\gamma$-collider where the high energy photons are extracted from the electron beam with Compton back-scattering \cite{Telnov:1989sd}. The $\gamma\gamma$-collider is considered as a future option with the integrated luminosity of the order of $100\fb^{-1}$ yearly \cite{Behnke:2013xla, Behnke:2013lya}. The machine could be upgraded to $\sqrt{s}=1\tev$ with the total integrated luminosity up to $300\fb^{-1}$ yearly. It should be noted that the CEPC and the FCC-ee are circular colliders, and a $\gamma\gamma$-collider could only be hosted on a linear collider, the ILC or the CLIC projects. Additionally, a $\gamma\gamma$-collider will provide a distinct way to produce the chargino pair which deserves a detailed study.

    The neutralino and the chargino pairs in $e^+e^-$-colliders were studied before at the next-to-leading order (NLO) accuracy by including the infrared divergences \cite{Oller:2005xg, Diaz:2009um, Zhu:2004ei, Heinemeyer:2017izw}. The chargino pair production rates in a $\gamma\gamma$-collider were also inspected before at the tree-level \cite{Mayer:2003sw, Klamke:2005jj} and at the loop-level \cite{Zhou:1999wx} without treating the infrared corrections. 
    In this work, the numerical calculation for the chargino pair production including all the possible one-loop level Feynman diagrams and also the radiative corrections are presented in a $\gamma\gamma$-collider. The numerical calculation is performed for the benchmark points that were proposed after the results obtained at the LHC with $\sqrt{s}=7, 8\tev$ collision data \cite{Baer:2013ula}. They are defined as follows:
        \begin{enumerate}
        \item[i.)]      Radiatively driven natural Susy (RNS), 
        \item[ii.)]     Natural Susy and mSUGRA/CMSSM scenario, 
        \item[iii.)]    Brummer Buchmuller (BB) benchmark point, 
        \item[iv.)]     Non-universal gaugino masses (NUGM) scenario.
        \end{enumerate}
    It should be emphasized that these scenarios and benchmark points are outside of the limits obtained by the LHC so far because the masses of the squarks and the sleptons are specifically arranged to be above the TeV scale. The masses of the lightest neutralino and the chargino are deliberately set below TeV range which makes them accessible at the Future Lepton Colliders (FLC) particularly in $\gamma\gamma$-collisions. The potentials of the FLCs and the $\gamma\gamma$-collider can be seen in these benchmark models about the supersymmetry searches. 
    In this study, the cross section of the lightest chargino pair is calculated as a function of the center-of-mass (c.m.) energy. Also, the distributions for the total cross section integrated with the photon luminosity in a $e^+e^-$-collider are discussed. 
    %\rred{add some more about what has been done in this study}.

    The content is organized as follows: In Sec. \ref{sec2}, the chargino sector in the MSSM is given briefly.  In Sec. \ref{sec3}, the one loop Feynman diagrams, the radiative corrections, and the convolution of the cross section in a $e^+e^-$ machine are presented. The stabilization of infrared and ultraviolet divergences are discussed. In Sec. \ref{sec4}, numerical results of the total cross section for each of the benchmark points are delivered. At last, the conclusion is drawn in Sec. \ref{sec5}.

%%%%%%%%%%%%%%%%%%%%%%%%%%%%%%%%%%%%%%%%%%%%%%%%%%%%%%%%%%%%%%%%%%%%%%%%
%%%%%%%%%%%%%%%%%%%%%%%%%%%%%%%%%%%%%%%%%%%%%%%%%%%%%%%%%%%%%%%%%%%%%%%%

\section{Expressions for the chargino sector}
\label{sec2}

    In Susy, each SM particle has a supersymmetric counterpart. The superpartners of W-boson and the charged Higgs bosons are Wino and Higgsinos, respectively. Charginos are the mass eigenstates of these particles, and they are defined as a linear combination of the Wino and the charged Higgsinos. The relevant part of the MSSM Lagrangian, that is responsible for the chargino masses, is defined as follows: 
        \begin{equation}
        \mathcal{L}=-\frac{1}{2}
        \begin{pmatrix}  \psi_i^+ & \psi_i^- \end{pmatrix} 
        \begin{pmatrix}  0 & \mathcal{M}_{\tilde\chi^\pm}^T \\   \mathcal{M}_{\tilde\chi^\pm} & 0 \end{pmatrix} 
        \begin{pmatrix}  \psi_i^+ \\ \psi_i^- \end{pmatrix}.
        \end{equation}
    The mass eigenstates in Dirac notation $\widetilde{\chi}_i^{\pm}$ are obtained using the Weyl states $\psi_i^+=(-i\lambda^+, \psi^1_{H_2})$ and $\psi_i^-=(-i\lambda^-, \psi^2_{H_1})$ with the following relations:
    \begin{equation}
        \widetilde{\chi}_i^{\pm}= \left( \begin{array}{ll} {\chi}_i^{\pm} \\ \bar{\chi}_i^{\mp}\end{array}  \right), 
        \;\;\;\; {\chi}_i^+=V_{ij}\psi_j^+,
        \;\;\;\; {\chi}_i^-=U_{ij}\psi_j^-.
    \end{equation}
   The mass matrix of the charginos is given below:
    \begin{equation}
     \mathcal{M}_{\tilde\chi^\pm} =  \left(
        \begin{array}{ll}
            M_{SU(2)}            & \sqrt{2}m_W\cos\beta        \\
        \sqrt{2}m_W\sin\beta     & \mu 
        \end{array} 
        \right)
        \label{eq:neumix}
    \end{equation}
    where $m_W$ is the mass of the W-boson, $\tan\beta=v_2/v_1$ is the ratio of the vacuum expectation values of the Higgs fields, $M_{SU(2)}$ represents the gaugino mass parameter associated with the $SU(2)$ symmetry group, and $\mu$ is the supersymmetric Higgs mass parameter. The diagonalization of mass matrix of the charginos $\mathcal{M}_{\tilde\chi^\pm}$ is performed with two unitary matrices $U$ and $V$.
    \begin{equation}
        diag(\chip{1}, \chip{2})= U^*\mathcal{M}_{\tilde\chi^\pm}V^\dagger.
    \end{equation}
    In general, the mass parameters in the chargino mass matrix could be complex for CP non-invariant cases. Since the CP violation is ignored in this study, all the parameters in the mass matrix are taken as real.

%%%%%%%%%%%%%%%%%%%%%%%%%%%%%%%%%%%%%%%%%%%%%%%%%%%%%%%%%%%%%%%%%%%%%%%%%
%%%%%%%%%%%%%%%%%%%%%%%%%%%%%%%%%%%%%%%%%%%%%%%%%%%%%%%%%%%%%%%%%%%%%%%%%
\section{The calculation of the diagrams}
\label{sec3}

    In this section, the analytical expressions related to the cross section, and the convolution process of the chargino pair production with the photon luminosity in $e^+e^-$-collider are provided. The scattering process is denoted as follows:
    \begin{eqnarray*}
    \gamma (k_1,\mu)+\;\gamma(k_2,\nu)\;\rightarrow \;  \tilde{\chi}^+_i (k_3)+\;\tilde{\chi}^-_j(k_4) \;\;\; (i, j=1,2)\,,\nonumber
    \end{eqnarray*}
    where $k_a$ $(a=1,...,4)$ are the four momenta of the incoming photons and outgoing charginos, also, $\mu$ and $\nu$ represents the polarization vectors of the incoming photons.

%%%%%%%%%%%%%%%%%%%%%%%%%%%%%%%%%%%%%%%%%%%%%%%%%%%%%%%%%%%%%%%%%%%%%%%%%
\subsection{The contributing Feynman diagrams}

\begin{figure}[htbp]
        \centering
            \includegraphics[width=0.40\textwidth]{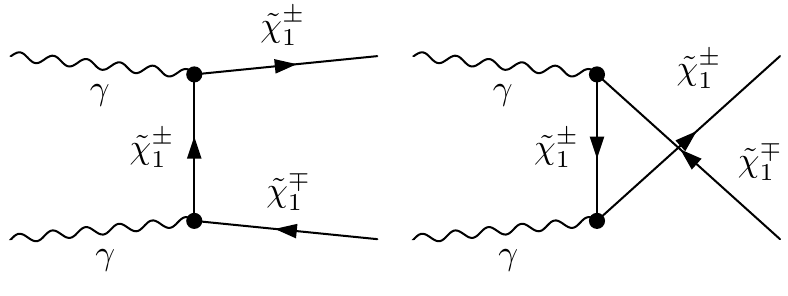}
        \caption{\label{fig1}    Tree level Feynman diagrams for the chargino pair production via photon-photon fusion.}
        \end{figure}

    In Fig. \ref{fig1}, the tree level Feynman diagrams contributing to the process $\gamma \gamma \rightarrow \tilde{\chi}^+_i \tilde{\chi}^-_j$ are plotted. Contrary to the neutralino pair production ($\gamma\gamma\rightarrow \tilde{\chi}_1^0\tilde{\chi}_1^0$) given in Ref. \cite{Sonmez:2015aa}, the chargino pair production is possible at the tree level. The diagrams are generated, then the amplitudes are constructed using \textsc{FeynArts}\cite{Hahn:2000kx}. The relevant part of the Lagrangian and the corresponding vertices are defined in Ref. \cite{Haber:1984rc}, and the \textsc{FeynArts} implementation of these rules are given in Ref. \cite{Hahn:2001rv}. Next, the total amplitude is handed to \textsc{FormCalc} \cite{Hahn:2000jm, Hahn:2010zi} for further evaluation.
    After summing over the helicities of the charginos and averaging over the polarization vectors of the incoming photons, the cross section of the unpolarized photon-photon collisions at the tree level is calculated by 
        \begin{equation}
        \hat{\sigma}_{\text{tree}}(\hat{s})=\frac{ \lambda( \hat{s},m_{\tilde\chi_1^\pm}^2 )}{16 \pi \hat{s}^2} \left(\frac{1}{4} \sum_{hel}{|\mathcal{M}_\text{tree}(\mu,\nu)|^2}\right)\,,
        \label{eq:partcross}
        \end{equation}
        where 
%    $\lambda( \hat{s},m_{\tilde\chi_i^+}^2, m_{\tilde\chi_j^-}^2 )=\sqrt{ (\hat{s}-m^2_{\tilde\chi_i^+}-m^2_{\tilde\chi_j^-})^2-4m^2_{\tilde\chi_i^+}m^2_{\tilde\chi_j^-} }/2$       
        $\lambda( \hat{s},m_{\tilde\chi_1^\pm}^2 )=\sqrt{ \hat{s}^2/4-\hat{s}\,m^2_{\tilde\chi_1^\pm} }$, the factor $\frac{1}{4}$ is the average of polarization vectors of the photons, $i$ and $j$ run over the flavor of the charginos at the final state, and finally $\mu,\;\nu$ represent the polarization of the photons. 

\begin{figure}[htbp]
    \centering
        \includegraphics[width=0.45\textwidth]{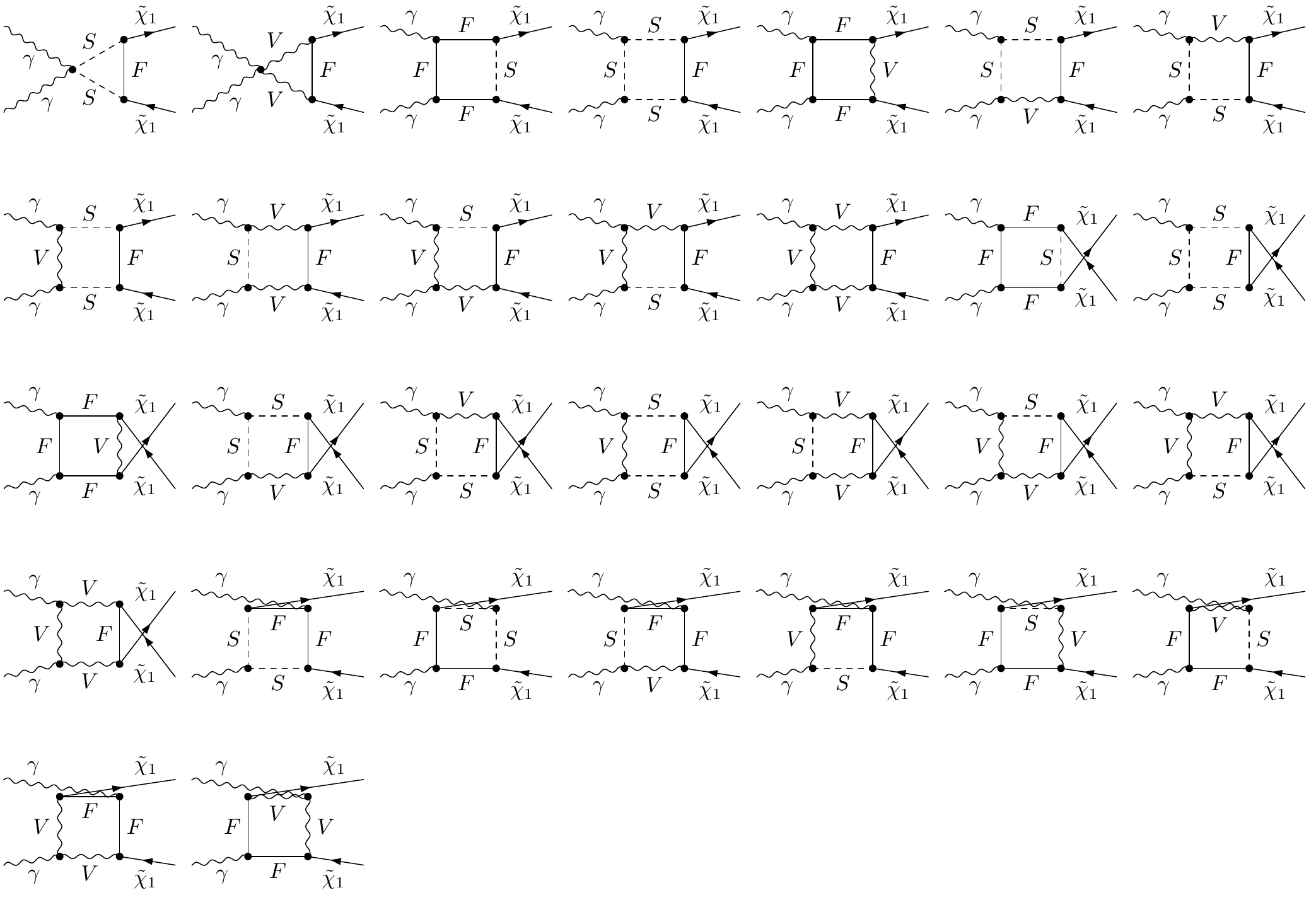}
    \caption{\label{fig2} One-loop Feynman box diagrams for the chargino pair production via photon-photon fusion.}
    \end{figure}

\begin{figure}[htbp]
    \centering    
        \includegraphics[width=0.45\textwidth]{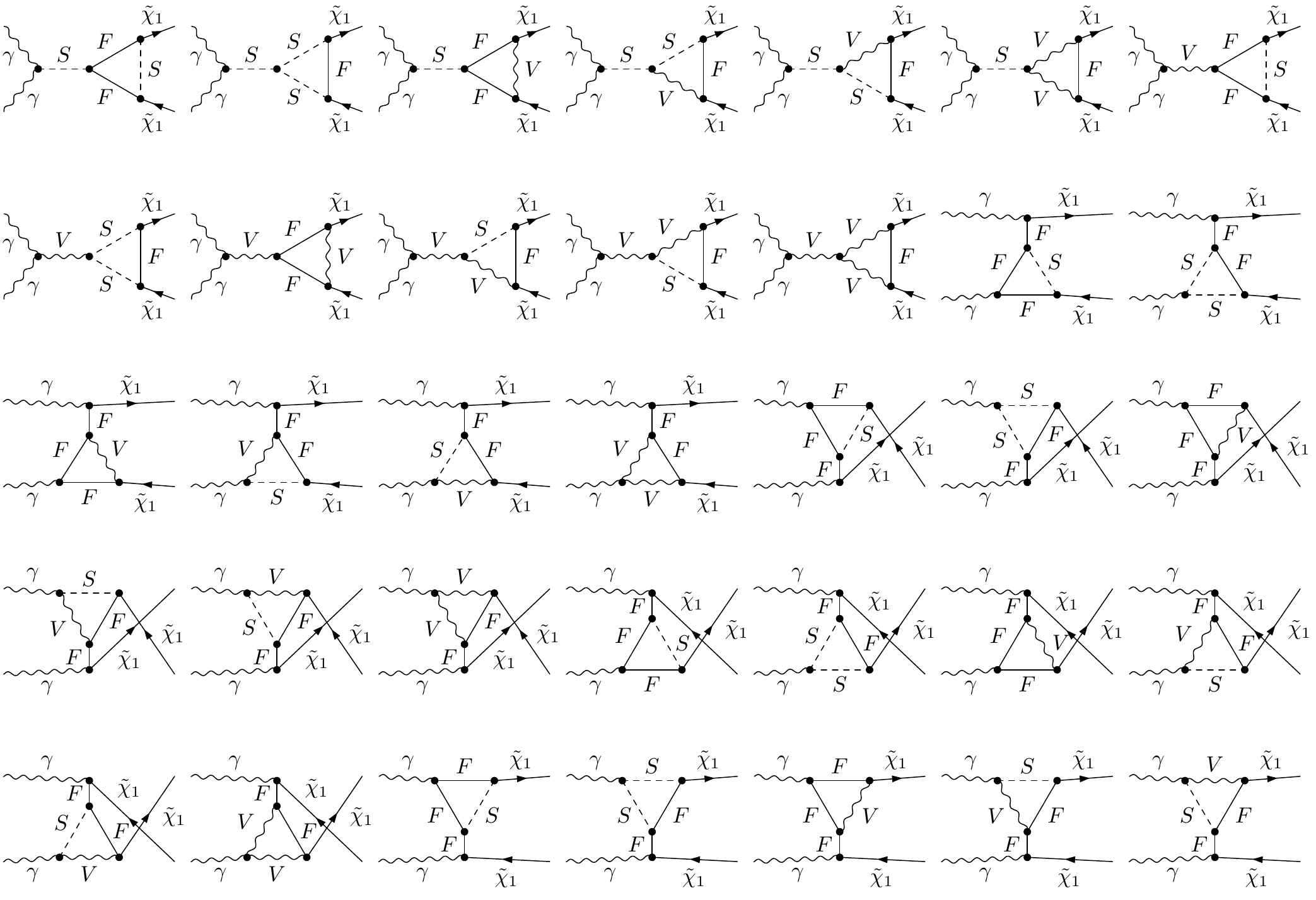}
        \includegraphics[width=0.45\textwidth]{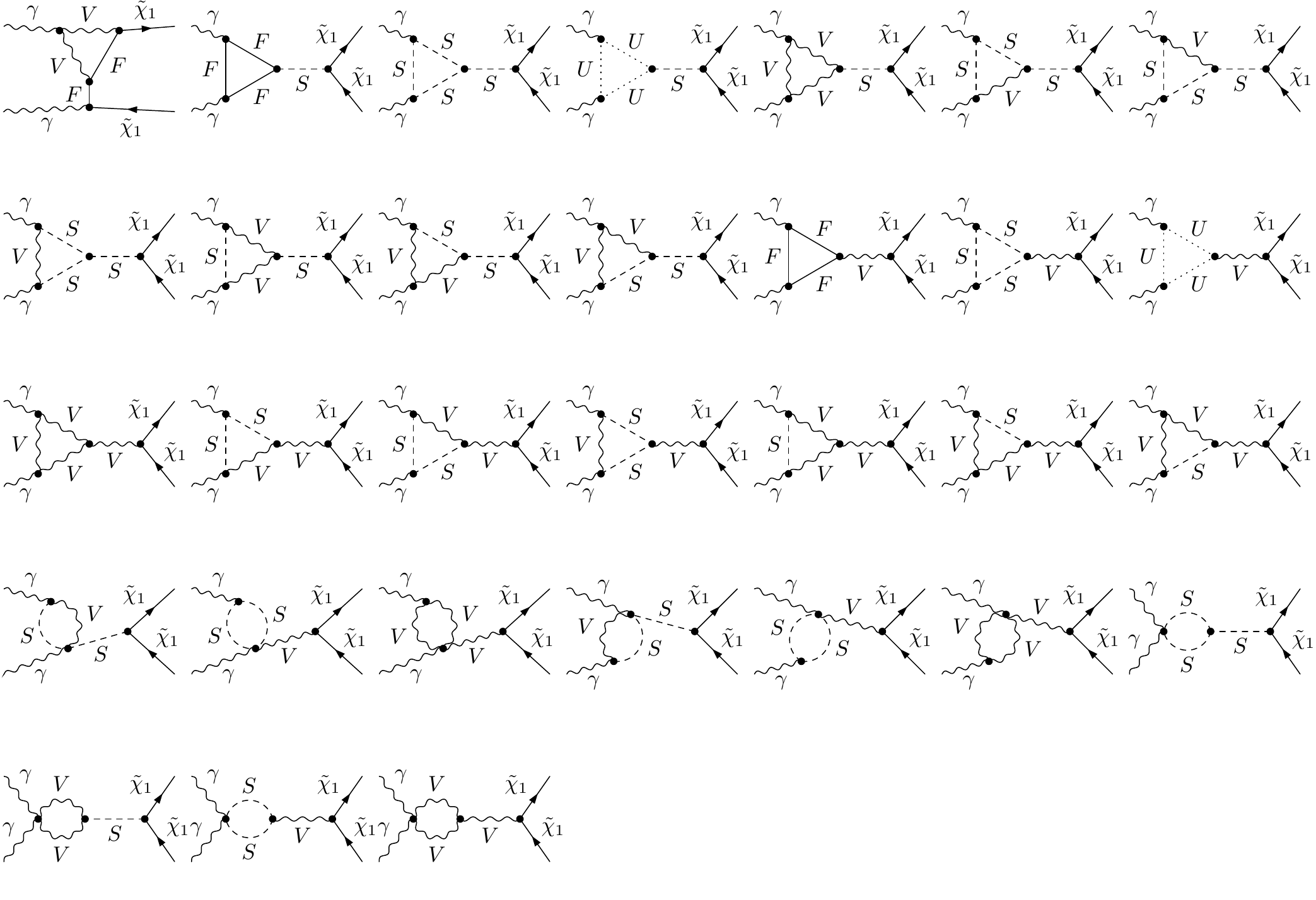}
    \caption{\label{fig3} Triangle- and bubble-type s-channel Feynman diagrams for the chargino pair production via photon-photon fusion.}
    \end{figure}

\begin{figure}[htbp]
    \centering    
        \includegraphics[width=0.48\textwidth]{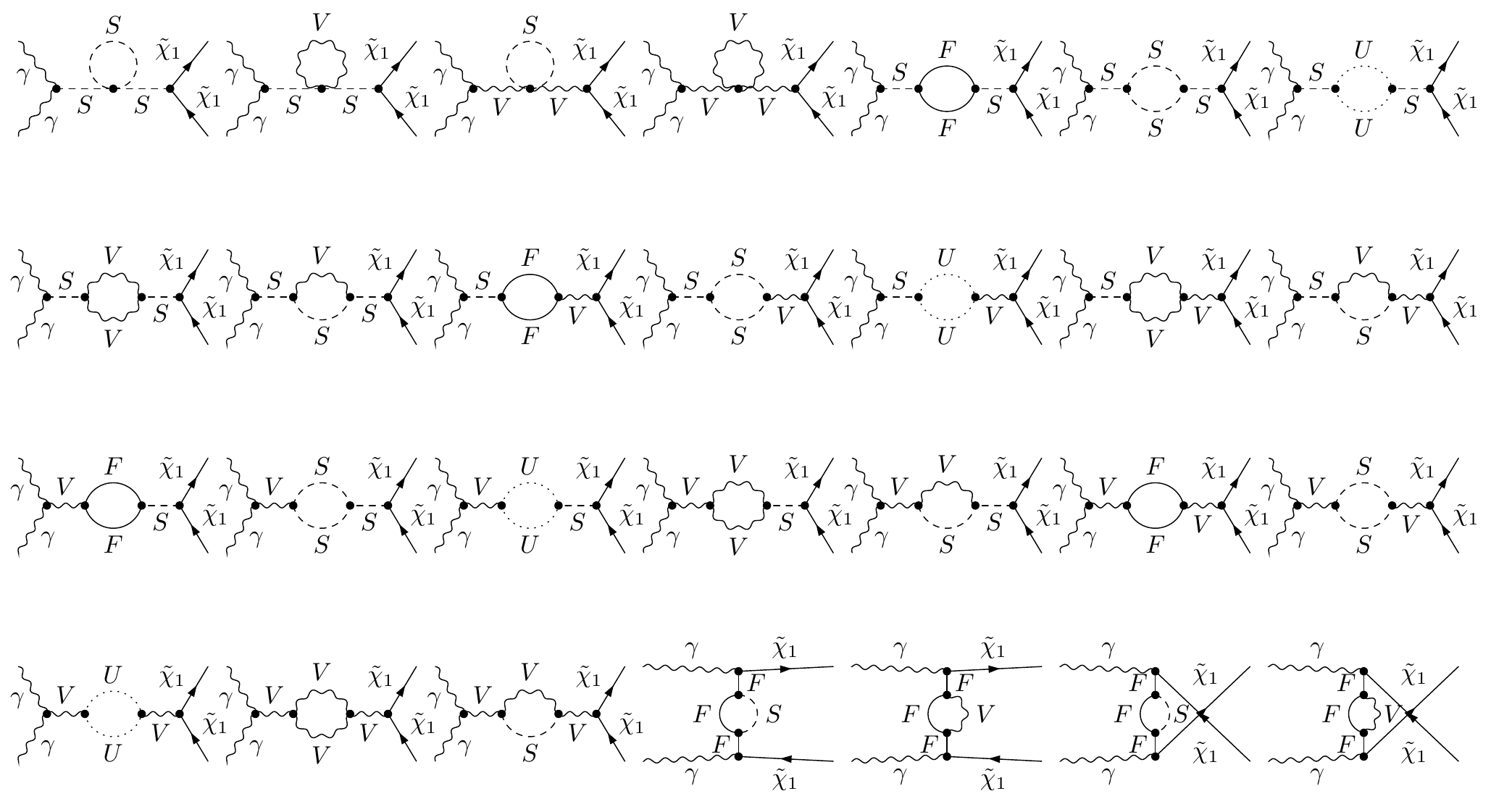}
    \caption{Feynman self-energy diagrams for the chargino pair production via photon-photon fusion.}
    \label{fig4}
    \end{figure}

    We classify the one-loop diagrams into three distinct groups, namely as box-type, triangle- and bubble-type s-channel diagrams, and self-energy diagrams. They are depicted in Fig. \ref{fig2}-\ref{fig4} where F states a fermion in the propagator, S for a scalar particle, and V for a vector boson propagator. They could be the SM particles or their supersymmetric counterparts. 

    The corresponding Lorentz invariant matrix element for the one-loop level process is written as a sum over the box-diagrams (Fig. \ref{fig2}), the triangle- and bubble-type diagrams (Fig. \ref{fig3}), and the self-energy diagrams (Fig. \ref{fig4}),
    \begin{equation}\label{eq6}
    {\cal M}_\text{virt}= {\cal M}_{box}+ {\cal M}_{tri}+ {\cal M}_{self}\,.
    \end{equation} 
    Next, the one-loop virtual correction is calculated by the following formula.
        \begin{equation}
        \hat{\sigma}_{\text{virt}}(\hat{s})=\frac{ \lambda( \hat{s},m_{\tilde\chi_i^+}^2, m_{\tilde\chi_j^-}^2 )}{16 \pi \hat{s}^2} \frac{1}{4} \sum_{hel} 2 Re \left[ {\mathcal{M}^*_\text{tree}\mathcal{M}_\text{virt}} \right]\,
        \label{eq:virtcross}
        \end{equation}
        where $\hat{s}$ represents the c.m. energy in the photon-photon collision frame.

%%%%%%%%%%%%%%%%%%%%%%%%%%%%%%%%%%%%%%%%%%%%%%%%%%%%%%%%%%%%%%%%%%%%%%%%%
\subsection{Ultraviolet and infrared divergences}

\begin{figure}[htbp]
        \centering
            \includegraphics[width=0.40\textwidth]{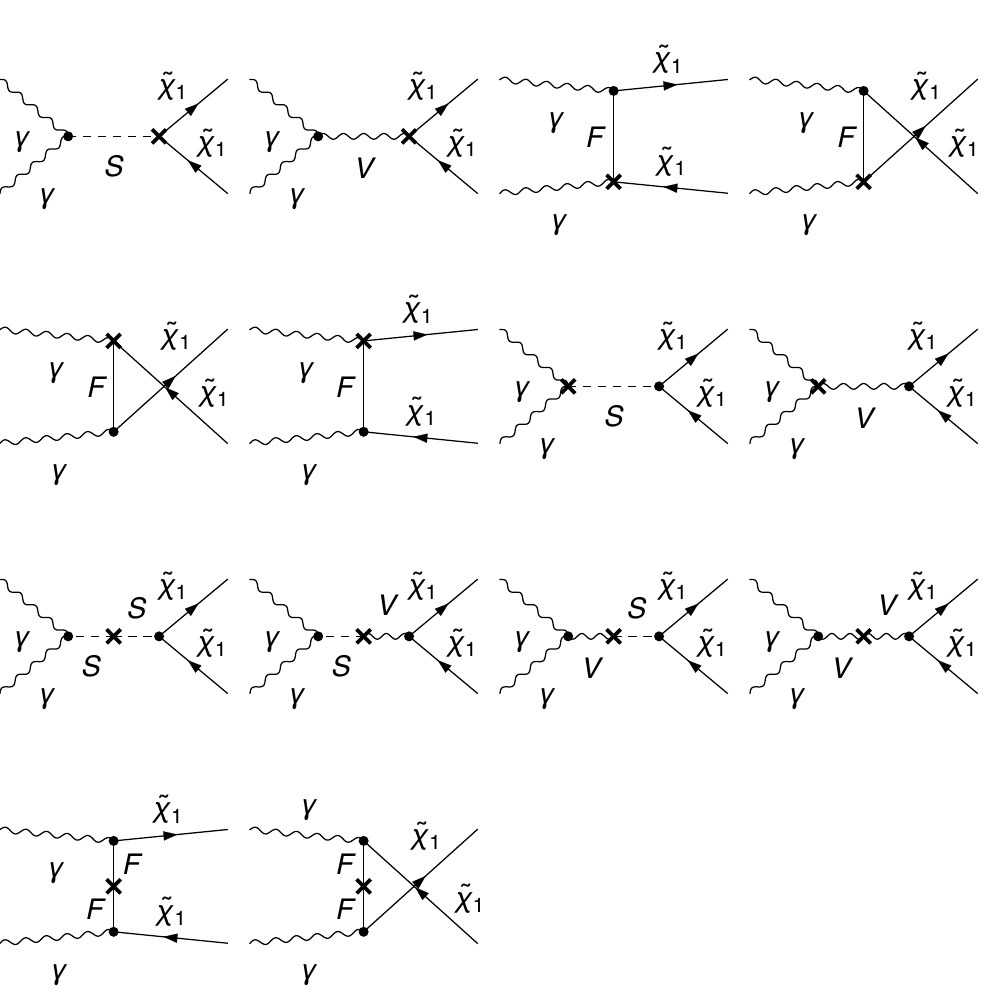}
        \caption{The Feynman diagrams that renormalize the relevant divergent vertices in the chargino pair production.}
        \label{fig5}
        \end{figure}

%\textbf{ultraviolet}\\
    In the computation, the ultraviolet divergence (UV) arising in the calculation is cured by taking into account the renormalization constants. The calculation is performed in the \emph{'t Hooft-Feynman} gauge because the gauge boson propagators are in simple form, hence the calculation requires less computing power. We used the constrained differential renormalization \cite{Siegel:1979wq} that is equivalent to the dimensional reduction \cite{delAguila:1998nd, Capper:1979ns} at the one-loop level \cite{Hahn:1998yk}. Besides, that also preserves the supersymmetry \cite{Stockinger:2005gx, Hollik:2005nn} and guarantees that the Susy relations are kept intact. All the counterterms and the renormalized vertices are indicated by a cross sign in Fig. \ref{fig5}. After the renormalization procedure, ${\cal M}_\text{virt}$ becomes UV finite.
    Nevertheless, the UV finiteness is checked numerically by varying the parameters $\mu$ and $\Delta$ on a large scale in \textsc{LoopTools} \cite{vanOldenborgh:1989wn, Hahn:2000jm, Hahn:2010zi}. These two parameters regularize the divergence that appeared in the scalar and the tensor one-loop integrals. The analysis showed that the cross section is stable in the numerical precision.

\begin{figure}[htbp]
        \centering
            \includegraphics[width=0.4\textwidth]{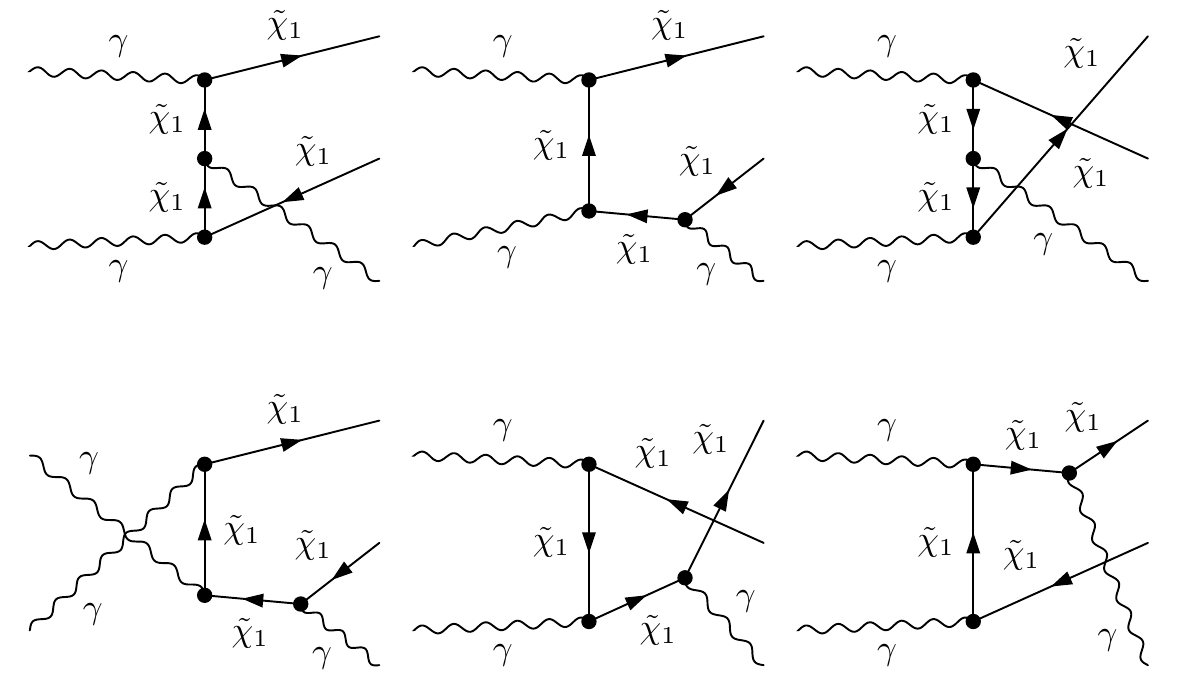}
        %\caption{The Feynman diagrams that contributes to the radiative corrections in the chargino pair production via photon-photon fusion. These diagrams show the process $\gamma\gamma\rightarrow\widetilde{\chi}_1^{+}\widetilde{\chi}_1^{-}\gamma$.}
        \caption{The relevant Feynman diagrams that contribute to the radiative corrections in chargino pair production. These diagrams show the process $\gamma\gamma\rightarrow\widetilde{\chi}_1^{+}\widetilde{\chi}_1^{-}\gamma$.}
        \label{fig6}
        \end{figure}

%\textbf{infrared}\\
    Besides the UV divergence, infrared divergence (IR) could also arise in the computation due to the massless particles propagating with very small energy in the diagrams. These diagrams lead to a singularity and what is called the IR-divergence occurs. If the photon had a mass such as $\lambda$, these divergent terms would be proportional to $\log \lambda$. This problem is cured by the fact that in nature an experimental apparatus performs measurement with a finite resolution and a minimum energy threshold. Therefore, if there are photons emitted with an energy of less than $\Delta E$, the apparatus actually could not measure them. The possibility of emitting soft photons has the same kind of IR singularity that appears in the loop calculations with the opposite sign. They need to be included in the computation accordingly. The diagrams plotted in Fig. \ref{fig6} have an additional photon at the final state, and adding these soft photon contributions cancels those IR-divergent ones that appeared in the loop diagrams with an internal photon.

\begin{figure}[htbp]
        \centering      
        \includegraphics[width=0.480\textwidth]{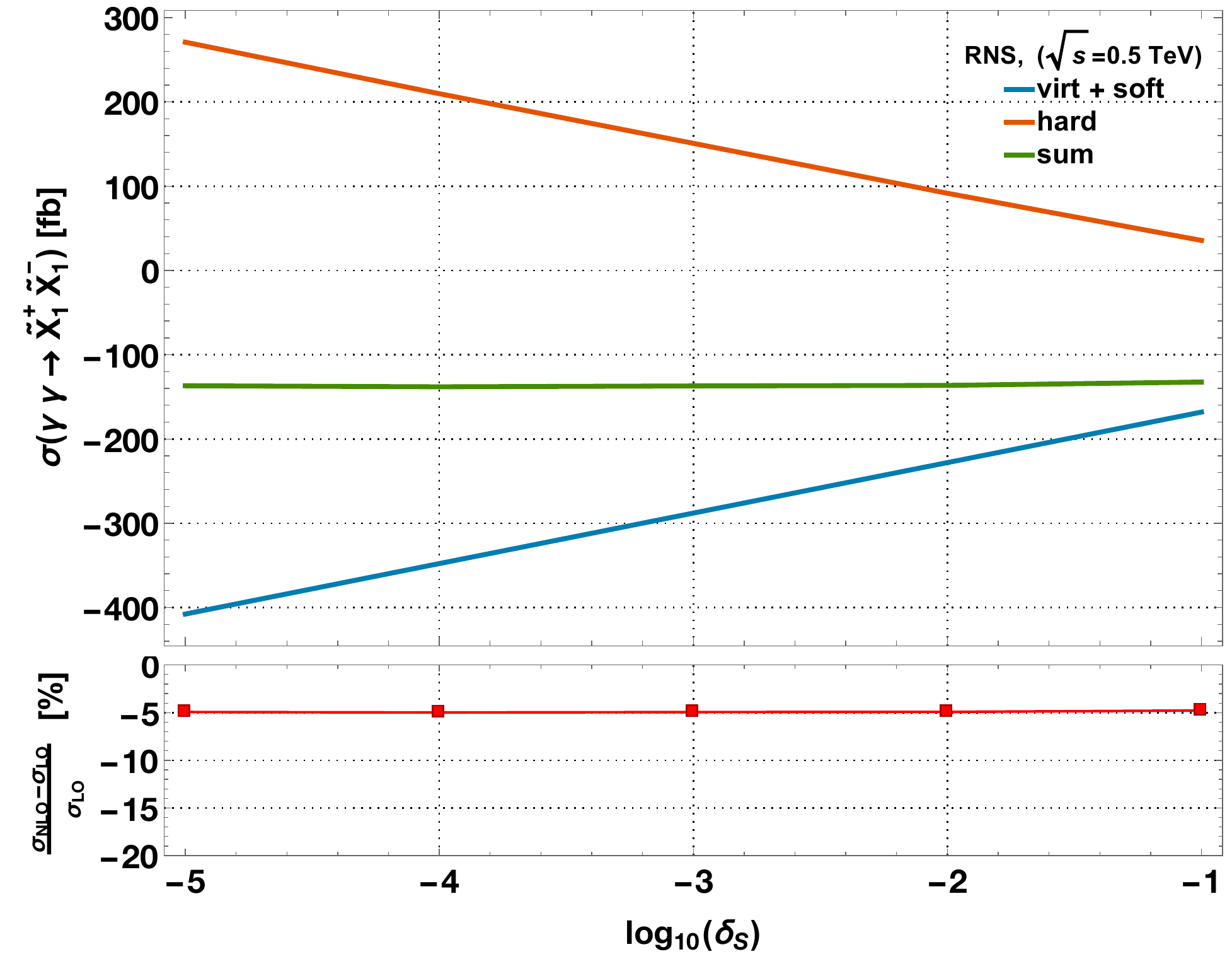}
        \includegraphics[width=0.480\textwidth]{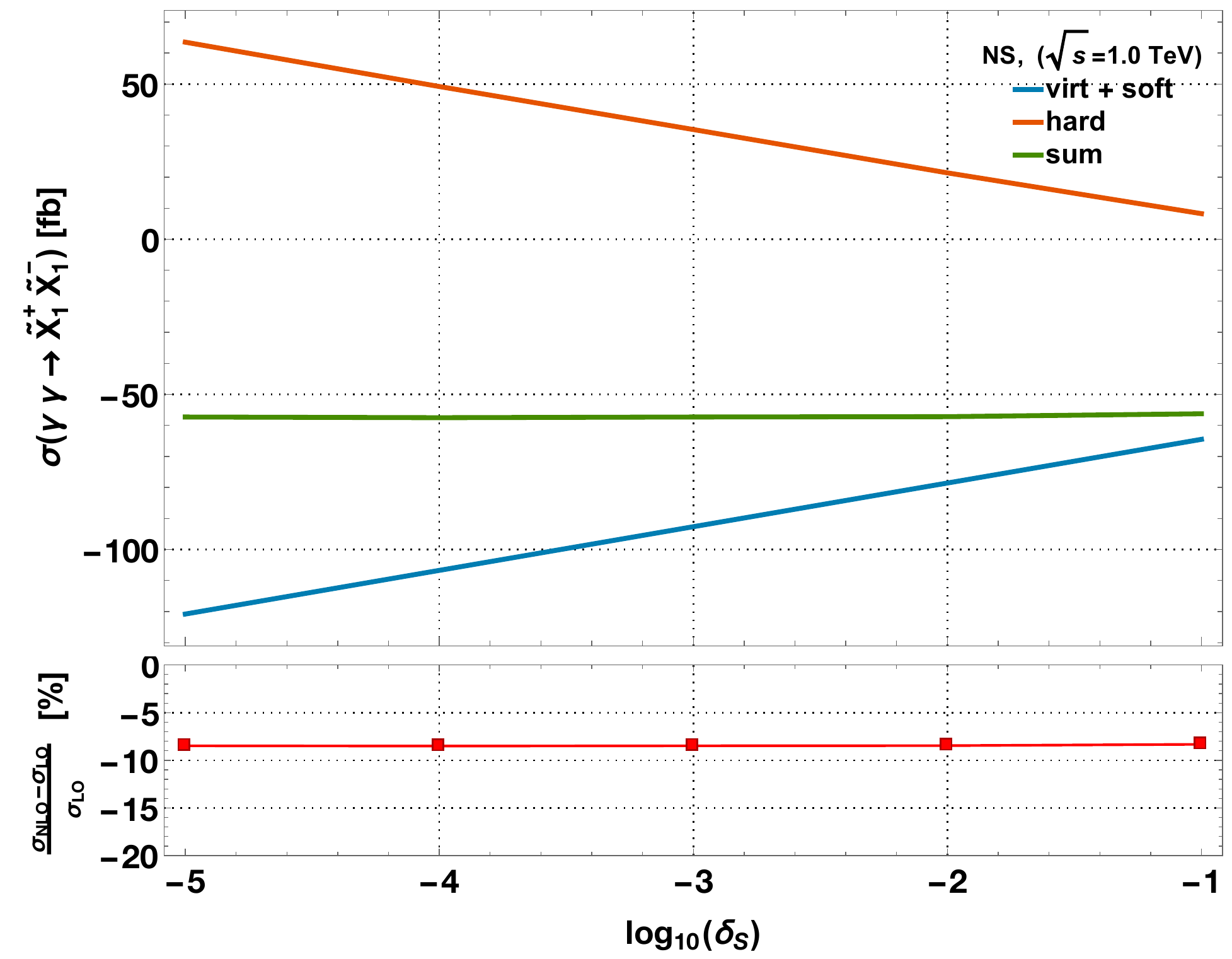}
        \caption{\label{fig7} The sum of the radiative corrections to the $\gamma\gamma\rightarrow\tilde\chi_1^+\tilde\chi_1^-$ as a function of $\delta_s = \Delta E/\sqrt{\hat{s}}/2$. The benchmark point RNS at $\sqrt{\hat{s}}=0.5\tev$ (NS at $\sqrt{\hat{s}}=1\tev$) is plotted at the left (right). The calculation is carried out for $\lambda=1$.}
        \end{figure}

    Fortunately, the soft bremsstrahlung correction is already implemented in \textsc{FormCalc} following the description given in Ref. \cite{Denner:1991kt}. The photons are considered soft, if their energy is less than $\Delta E =\delta_s E=\delta_s \sqrt{\hat{s}}/2$ that separates soft and hard photon radiation. Since the diagrams with a photon propagating in the loops are regularized by the photon mass parameter $\lambda$, adding the virtual ($\sigma_{\text{virt}}(\lambda)$) and the soft ($\sigma_{\text{soft}}(\lambda,\Delta E)$) contributions drops that dependence. % on the photon mass parameter $\lambda$. 
    On the other hand, the contribution coming from the hard photon radiation needs to be combined to obtain a complete picture of the process. Otherwise, the total cross section will be dependent on the $\Delta E$. In the computation, the cancellation of these IR divergences and the dependence on the parameter $\lambda$ are investigated. The sum of the virtual and the soft photon radiation is stable in varying the parameter $\lambda$ on a large scale. This stability shows that the dependence on the $\lambda$ is dropped out. Next, the contribution coming from the hard photon emission is also computed as a function of $\delta_s$. In Fig. \ref{fig7}, all the contributions are plotted for the benchmark point RNS (NS) at the left (right). The total NLO correction is around $-8.3\%$ for the NS ($\sqrt{\hat{s}}=0.5\tev$) and $-4.8\%$ for the RNS ($\sqrt{\hat{s}}=1\tev$). The sum (green line) for both of the distributions are stable by varying the $\delta_s$ logarithmically. The stability is also seen at varying the c.m. energy. In conclusion, the total one-loop corrections are decomposed into the virtual, the soft, and the hard parts as follows:
        \begin{eqnarray}
            \hat{\sigma}^{\text{NLO}}&=&\hat{\sigma}_{\text{virt}}(\lambda)+\hat{\sigma}_{\text{soft}}(\lambda,\Delta E)+\hat{\sigma}_{\text{hard}}(\Delta E).
                            %&&+\sigma_{\text{hard}}(\Delta E,\delta_\theta)+\sigma_{\text{coll}}(\Delta E,\delta_\theta).
        \end{eqnarray}

%%%%%%%%%%%%%%%%%%%%%%%%%%%%%%%%%%%%%%%%%%%%%%%%%%%%%%%%%%%%%%%%%%%%%%%%%

\subsection{Convoluting the cross section with the photon luminosities}
    
    The production of chargino pair could also be studied in a future $e^+e^-$-collider where the photon beam is extracted by Compton back-scattering technique. The big fraction of the c.m. energy of the electron beam could be transferred to the photons. Then, the production of the chargino pair could be taken as a sub-process in $e^+e^-$ collisions. The full process will be considered as $e^+e^-\rightarrow\gamma\gamma\rightarrow \tilde\chi_1^+\tilde\chi_1^-$, and the total cross section could be computed by convoluting the cross section $\hat{\sigma}_{\gamma\gamma\rightarrow \tilde\chi_1^+\tilde\chi_1^-}(\hat{s})$ with the photon luminosity in a $e^+e^-$-collider.
    \begin{eqnarray}
    \label{eq:total_cross}
    \sigma(s)=\frac{1}{4} \sum_{\alpha,\beta=\pm1} \int_{4m_{\tilde\chi_1^\pm}^2/s}^{y_m^2} d\tau \frac{dL_{\gamma\gamma}}{d\tau} 
    \left(1+\alpha\xi(y)\right) \left(1+\beta\xi(\tau/y)\right)\hat{\sigma}_{\alpha\beta} ,
    \end{eqnarray}   
%    \begin{equation}
%    \label{eq:total_cross}
%    \sigma(s)=\int_{4m_{\tilde\chi_1^\pm}^2/s}^{y_m^2} d\tau\, \frac{dL_{\gamma\gamma}}{d\tau} 
%    \left\{\frac{(1+\xi_1 \xi_2)}{2} \hat{\sigma}_{RR}+\frac{(1-\xi_1 \xi_2)}{2}\hat{\sigma}_{RL}\right\},
%    \end{equation}
    where $\hat{\sigma}_{\alpha\beta}=\hat{\sigma}_{\gamma^\alpha\gamma^\beta\rightarrow  \tilde\chi_1^+\tilde\chi_1^-}( \hat{s})$ is the scattering cross section for the polarization configurations of the incoming photon beams, $\hat{s}=\tau s$, and $(\alpha, \beta)$ represent the polarization of the photons. The $s$ and the $\hat{s}$ are the c.m. energy in $e^+e^-$ collisions and $\gamma\gamma$ sub-process, respectively. 
    %$\tau_{min}$ is the threshold energy to produce the chargino pair, and it is defined as $\tau_{min}=$. 
    The energy spectrum $F_{\gamma}(x,y)$ and the mean polarization $\xi(y)$ of the scattered photons are defined in Refs. \cite{Ginzburg:1982bs, Ginzburg_1984, Telnov:1989sd, Telnov:1995hc}, where $y=E_\gamma/E_e$ with $E_\gamma$ and $E_e$ being the energy of photon and electron beams, respectively. The energy spectrum of the photons includes only the Compton back-scattered photons, and nonlinear effects are not taken into account.
    %\emph{The nonlinear effects in the energy spectrum of the photons such as high density of the laser beam, the correlation between photon energy and the scattering angle (Compton back-scattering), electron re-scattering and two-photon scattering of the electrons are included.}
    The maximum fraction of the photon energy is defined as $y_m=x/(1+x)$ where $x=\left(4 E_e  E_l /m_e^2\right)$, the laser photon energy $E_l=1.17\text{ eV}$,
    %the collision angle $\alpha_0=2\cdot 10^{-6}$, 
    and $m_e$ is the electron mass. In this study, we set $x=4.8$. 
    In Refs. \cite{Ginzburg_1984, Telnov:1995hc}, it is stated that the photon luminosity also depends on the variable $\rho^2$ which defines the focal point between the conversion point and the interaction point; however, it depends on the instrumentation of the collider. Since the calculation is carried out for the assessment of the potential of $\gamma\gamma$-collider generally, $\rho^2=0$ is employed in the convolution.
    The photon luminosity is defined as follows:
    \begin{equation}
    \frac{dL_{\gamma\gamma}}{d\tau}=\int_{\tau/y_{m}}^{y_{m}}\frac{dy}{y}F_{\gamma}(x,y)F_{\gamma}\left(x,\tau/y\right)\,.
    \end{equation}

%%%%%%%%%%%%%%%%%%%%%%%%%%%%%%%%%%%%%%%%%%%%%%%%%%%%%%%%%%%%%%%%%%%%%%%%%
%%%%%%%%%%%%%%%%%%%%%%%%%%%%%%%%%%%%%%%%%%%%%%%%%%%%%%%%%%%%%%%%%%%%%%%%%
\section{Numerical Results and Discussion}
\label{sec4}
    
    In the analysis, the following input parameters are taken from Ref. \cite{Eidelman:2004wy} where $m_W = 80.399\gev$ $m_Z=91.1887\gev$, $m_t=173.21\gev$, $s_W^2=0.222897$, and $\alpha(m_Z)=1/127.944$. The prominent feature of the supersymmetry is that all the three gauge-couplings unify at the grand scale which is also predicted by GUTs and string theories. However, there are no any superpartners discovered at the weak scale. To accommodate this fact, it was assumed that the supersymmetry is slightly broken and all the superpartners acquire mass higher than the electroweak scale. Nevertheless, the breaking scale could not be at the order of ten TeV because the soft Susy breaking parameters are intimately linked to the breakdown of the electroweak symmetry. Accordingly, it was assumed that the masses of these sparticles are not far away from the electroweak scale. However, the results coming from the LHC at $\sqrt{s}=7,8,13\tev$ had made that lovely picture to fade away and questioned the simple weak scale Susy picture. Since the exclusion of the difference between the sparticle masses and the weak scale increases the breaking, unfortunately, that also resurrects the so-called little hierarchy problem \cite{Martin:1997ns}. 
    The benchmark points considered in this paper have been introduced to fit into this picture drawn by the LHC results, and they are chosen for having a low contribution to the electroweak observables. They were introduced explicitly for a future lepton-lepton collider by the constraints set from the LHC results. A detailed discussion about these benchmark points was delivered in \cite{Baer:2013ula} and references therein. To achieve small $\Delta_{EW}$, it is required that $|m^2_{H_u}|$, Higgs-doublet mixing parameter $\mu$, and the radiative contribution $|\Sigma_u^u|$ to be around $m^2_Z/2$ to within a factor of a few \cite{Baer:2012up, Baer:2012cf}. The results are presented for each benchmark point in the following subsections. The decay widths of the neutral Higgs bosons are obtained at the NLO-level accuracy employing \textsc{FeynHiggs}\cite{Hahn:2009zz, Hahn:2015gaa}, and the finite width scheme is introduced for them. The input parameters in \textsc{FeynHiggs} are taken from table \ref{tab:rnsparam}. 
    
    It should be noted that the second chargino mass in all these scenarios is beyond the TeV except for the RNS and the mSUGRA; nevertheless, their masses are higher than $0.5\tev$. Therefore, the production of $\tilde\chi_1^+\tilde\chi_2^-$ and $\tilde\chi_2^+\tilde\chi_2^-$ requires much greater c.m. energy than the lightest chargino. 

\subsection{Radiatively driven natural Susy (RNS) scenario} 

    This scenario is motivated by minimizing $\Delta_{EW}$, and it also sustains the unification of the gauge couplings. If it is ensured that the Higgs-doublet mixing parameter $\mu$ is $\sim100-300\gev$, that causes a small negative value of $m_{Hu}^2$ at the weak scale and large mixing between top squarks. The mass spectrum is calculated for the parameters given in Tab. \ref{tab:rnsparam} using \textsc{ISASUGRA-v7.88}  \cite{Paige:2003mg}\footnote{The SLHA files for these benchmark points could also be obtained from \url{http://flc.desy.de/ilcphysics/research/susy}.}. The masses of the lightest two neutralinos and the lightest chargino are around the electroweak scale in the RNS. However, all the other sparticle masses are beyond the TeV \cite{Baer:2016wkz, Baer:2013aa, Baer:2013xua}.        
        \begin{figure}[htbp]
        \centering      
        \includegraphics[width=0.48\textwidth]{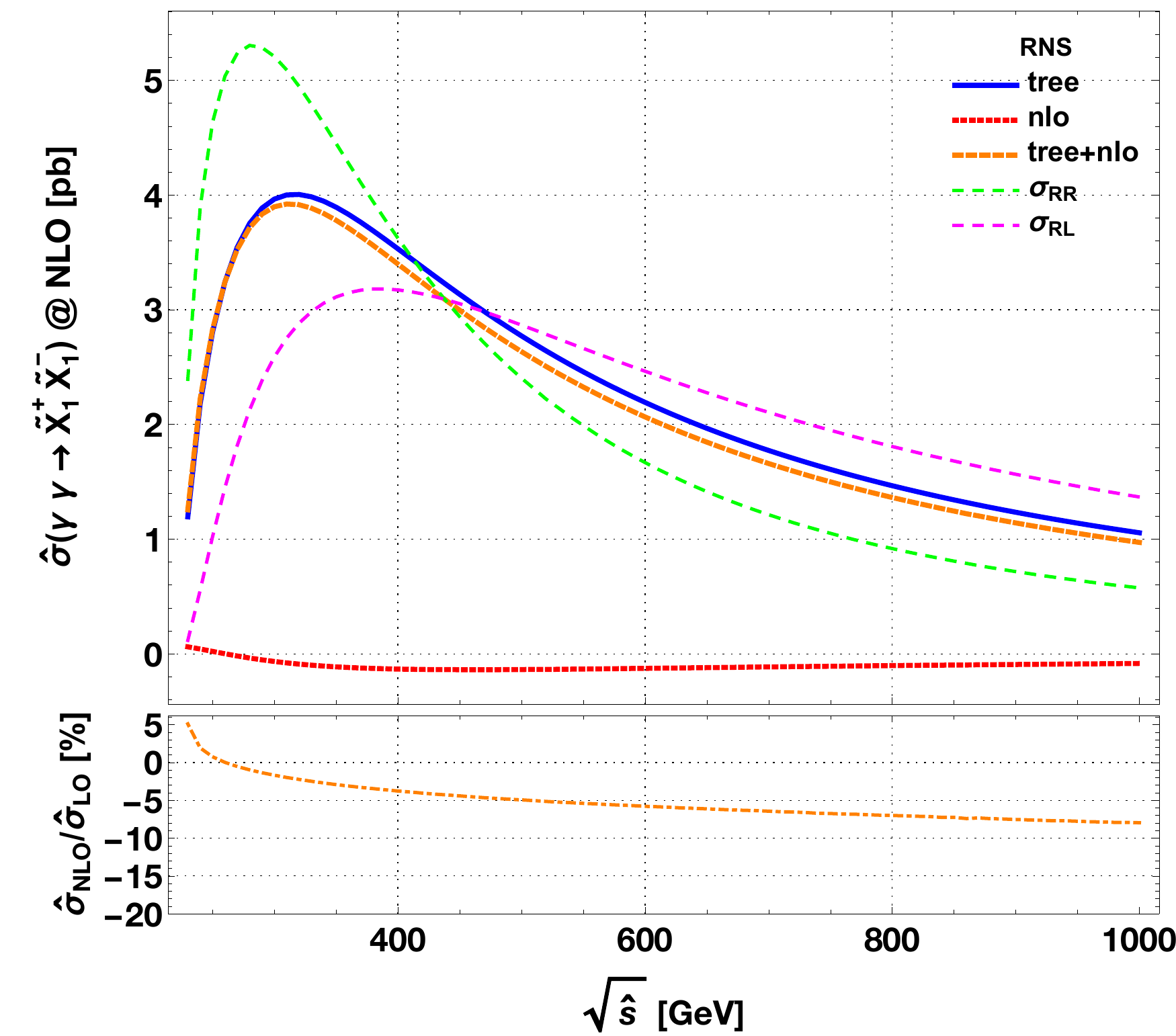}
        \caption{\label{fig8}
        The distribution of the cross section at the tree level (blue solid line), the total (virtual + real) NLO corrections (red dashed line), the sum of all (orange dotted line), and two polarization cases $\hat{\sigma}^\text{LO+NLO}_\text{RR}$ (green dotted line) and $\hat{\sigma}^\text{LO+NLO}_\text{RL}$ (magenta dotted line) are plotted for the process $\gamma\gamma\rightarrow\tilde\chi_1^+\tilde\chi_1^-$ as a function of $\hat{s}$. The ratio $\hat{\sigma}^\text{NLO}/\hat{\sigma}^\text{LO}$ is plotted as a function of the energy at the bottom.}
        \end{figure}

    The distribution of the cross section at the tree level, the total NLO corrections (virtual + real), and the sum of all are plotted in Fig. \ref{fig8} for the lightest chargino pair as a function of the c.m. energy. The ratio $\hat{\sigma}_\text{NLO}/\hat{\sigma}_\text{LO}$ is plotted at the bottom of the figure. It can be seen that the correction at the NLO level is positive in $\sqrt{\hat{s}}<250\gev$, and it falls quickly moving to higher c.m. energies. It reaches down to -7.96\% at $\sqrt{\hat{s}}\sim 1\tev$. 
    In Fig. \ref{fig8}, the unpolarized total cross section goes up to $3.94\pb$ around $\sqrtHatS=310\gev$, and it falls rapidly to $0.97\pb$ at $\sqrtHatS=1\tev$. The distributions of the $\hat{\sigma}^\text{LO+NLO}_\text{RR}$ ($J_z=0$) and the $\hat{\sigma}^\text{LO+NLO}_\text{RL}$ ($J_z=2$) are also given in Fig. \ref{fig8}. The total cross section is dominated by $J_z=0$ for $\sqrtHatS<440\gev$, and $J_z=2$ becomes greater in the region of $\sqrtHatS>440\gev$. This trend is the same for all the other scenarios analyzed in this paper, and indeed the contribution coming from the $J_z=0$ is greater at low c.m. energies, and the $J_z=2$ becomes higher at greater c.m. energies. 
    Moving to the second possible case for the chargino pair production, $\tilde\chi_1^+\tilde\chi_2^-$ requires collision energy greater than $\sim 725\gev$. Considering the energy of the back-scattered photons at $\sqrt{s}=1\tev$, the c.m. energy is just around the threshold, and the production cross section is less than ab. Therefore, the production of $\tilde\chi_1^+\tilde\chi_2^-$ and $\tilde\chi_2^+\tilde\chi_2^-$ pairs are not investigated.
            
    The branching ratios of the chargino are also calculated using \textsc{ISASUGRA}, and the possible pattern for the discovery of the lightest chargino is examined. In the RNS, the lightest chargino decays primarily through three channels, and the sum of these branching ratios is $\sim 0.999$. These channels are     
        $BR(\tilde\chi_1^+ \rightarrow \tilde\chi_1^0 + u + \bar{d}) = 0.333 $, 
        $BR(\tilde\chi_1^+ \rightarrow \tilde\chi_1^0 + c + \bar{s}) = 0.333 $, and 
        $BR(\tilde\chi_1^+ \rightarrow   \tilde\chi_1^0 + l + \nu_l ) = 0.333$. 
    Consequently, the pattern of a single chargino decay at the collider would be a missing transverse energy due to the LSP plus two light quark jets or a lepton. 
    %The second chargino decays through the lightest neutralino and either of $W^\pm$, $Z^0$, or $H^0$ boson. 
    %$\tilde\chi_2^+ \rightarrow  \tilde\chi_1^0  + W^+ = 0.238 $
    %$\tilde\chi_2^+ \rightarrow  \tilde\chi_2^0  + W^+ = 0.256 $
    %$\tilde\chi_2^+ \rightarrow  \tilde\chi_1^0  + Z^0 = 0.242 $ 
    %$\tilde\chi_2^+ \rightarrow  \tilde\chi_1^0  + H^0 = 0.241 $ 
        \begin{widetext}
        \onecolumngrid
        \begin{table}[htp]
        \caption{The input parameters and the mass spectrum for all the benchmark points considered in this study. All masses are in TeV. The mass spectrum and the electroweak scale parameters are obtained with \textsc{ISASUGRA}.}
        \begin{center}
%        \begin{ruledtabular}
        \begin{tabular}{m{1.7cm}|m{2cm}m{1.8cm}ccccp{2cm}c}
        Benchmark Point 
                & $m_0(1,2)$, $m_0(3)$ 
                                    & $m_{1/2}$, $M_{1,2,3}$    
                                                & $A_0$        & $\tan\beta$    
                                                                        & $\mu$    
                                                                                        & $m_{H^0/A^0/H^\pm}$  
                                                                                                        &  {$m_{\tilde{\chi}_{1/2}^\pm}$}  
                                                                                                                        &  {$m_{\tilde{\chi}_{1}^0}$} \\
        \hline
        RNS        &5                &    0.7        &    -8.3    &    10    &    0.11        &    1        &(0.113, 0.61)  & 0.101    \\
        NS        &13.35, 0.76
                                &    1.38    &   -0.167    &    23    &    0.225        &$\sim1.55$    &(0.233, 1.18)  & 0.224    \\
        mSUGRA    &10               &    0.5        &    -5.45    &    15    &    0.234        &$\sim9.70$    &(0.248, 0.70)  & 0.229    \\
        BB        & -                &    $\sim5.3$, $\sim9.5$    
                                            &    -        &    48        &    0.160        &$\sim4.05$    &(0.167, 9.52)  & 0.167    \\
        %NUHM2    &10               &    0.8        &    -16        &    7        &    0.280        &$\sim0.28$    &(0.237, 5.9)   & 0.237    \\
        NUGM    &3                &    0.7        &    -6        &    25        &    2.36        &$\sim3.30$    &(0.216, 2.36)    & 0.131 
        \end{tabular}
 %       \end{ruledtabular}
        \end{center}
        \label{tab:rnsparam}
        \end{table}%
        \end{widetext}

    %\textbf{angular distributions:}\\    
    %\rred{The angular distribution of each chargino pair for $0.5\,\tev$ and $1\,\tev$ c.m. energies are given in Fig. \ref{fig2} and \ref{fig3}. Where at low c.m. energies there is a small asymmetry at the same order for each neutralino pairs, therefore according to Fig. \ref{fig3} the asymmetry gets large for higher cms energy due to the asymmetrical $t-$ and $u-$terms presented in the cross section. The angular distribution of $\tilde{\chi}^0_1\tilde{\chi}^0_2$ is close to isotropy compared to the other chargino pairs.}        
                
\subsection{Natural Susy (NS) and mSUGRA/CMSSM scenarios} 

\begin{figure}[htbp]
        \centering
        \includegraphics[width=0.48\textwidth]{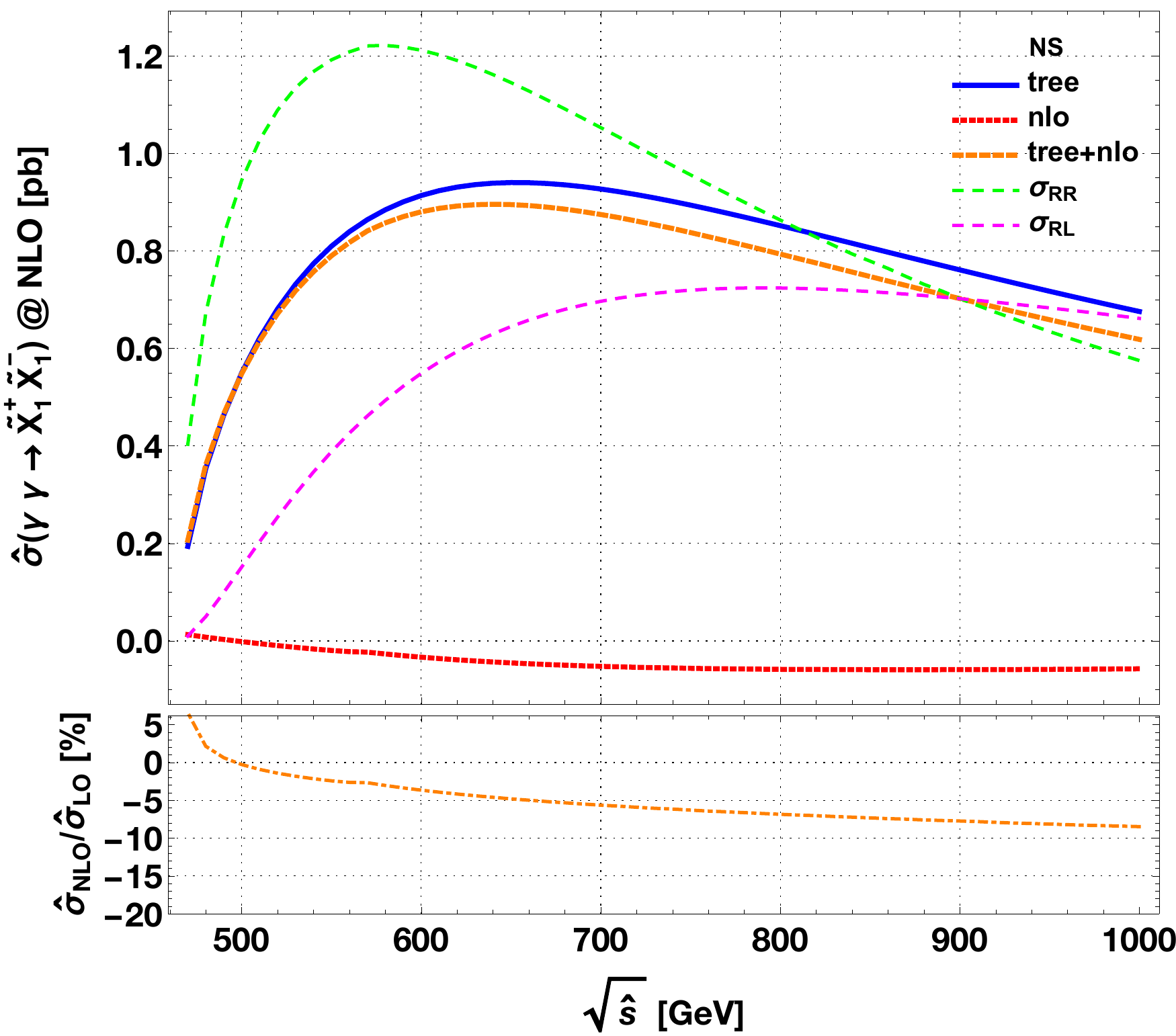}
        \includegraphics[width=0.48\textwidth]{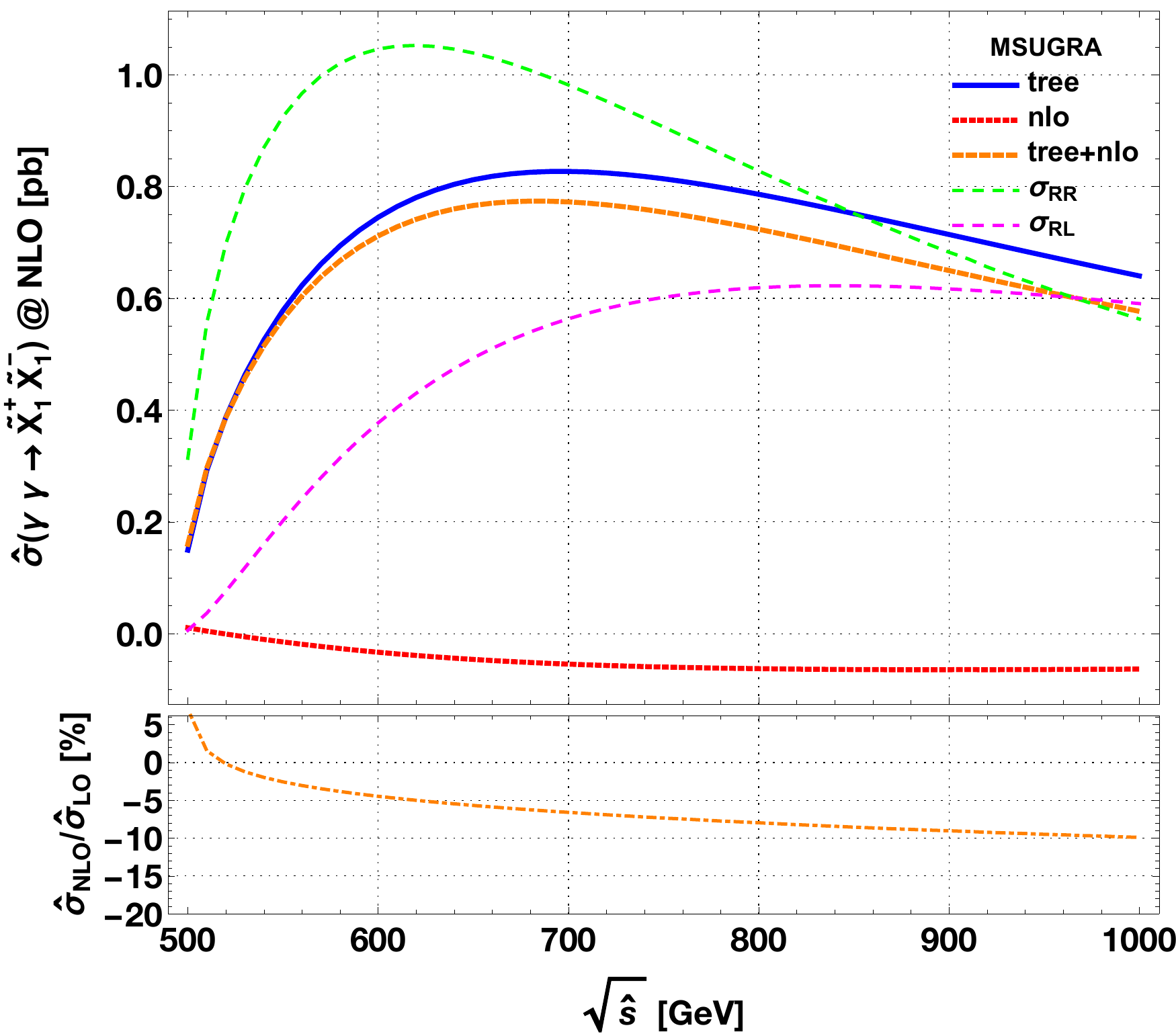}
        \caption{The distribution of the various cross sections of the process $\gamma\gamma\rightarrow\tilde\chi_1^+\tilde\chi_1^-$ as a function of $\sqrtHatS$. Two polarization cases $\hat{\sigma}^\text{LO+NLO}_\text{RR}$ and $\hat{\sigma}^\text{LO+NLO}_\text{RL}$ are indicated by green-dotted and magenta-dotted lines in the figure, respectively. The ratio $\hat{\sigma}^\text{NLO}/\hat{\sigma}^\text{LO}$ is plotted as a function of the energy at the bottom. (left): the distributions are for Natural Susy, (right): the mSUGRA benchmark scenario is assumed.  }
        \label{fig9}
        \end{figure}

%\textbf{Natural Susy Scenario}\\
    The Natural Susy (NS) was introduced and characterized in Refs. \cite{Papucci:2011wy, Feng:2013pwa, Brust:2011tb}. The lightest chargino mass for this benchmark point is $m_{\tilde{\chi}_{1}^{\pm}}\approx 233\gev$, and the second lightest one is greater than TeV. Besides of the third generation squarks ($\tilde{t}_{1,2}$ and $\tilde{b}_{1}$), all the other sparticles have a mass higher than TeV. The energy dependence of the total cross section (LO+NLO) of $\hat{\sigma}(\gamma\gamma\rightarrow\tilde{\chi}^+_1\tilde{\chi}^-_1)$ is plotted in Fig. \ref{fig9} (left). 
    The total cross section reaches up to $0.896\pb$ around $\sqrt{\hat s}=640\gev$ at the loop-level. 
    The total virtual correction is positive, and it reaches up to $+3\%$ for less than $\sim500\gev$. Then, it falls quickly, and it becomes negative for $\sqrtHatS>500\gev$. Overall, the virtual corrections lower the tree-level cross section up to 9\%. 
    
    Moving to the mSUGRA/CMSSM scenario, the LHC8 ruled out most of the region in the parameter space by direct searches of gluino and squarks. However, there still exists some space for the dark matter. This benchmark point employs the parameters at the GUT scale given in Tab. \ref{tab:rnsparam}, and the electroweak parameters are calculated by \textsc{ISASUGRA}. Since the Higgs-doublet mixing parameter $\mu$ is $\sim 235\gev$, the masses of the neutralino and the chargino (electroweakinos) are around the electroweak scale. Similar to the previous scenarios, all the other sparticles are over TeV. Therefore, this point is beyond the reach of the LHC. The production rate of the chargino pair for the unpolarized photon beam is given in Fig. \ref{fig9} (right) as a function of the c.m. energy. The unpolarized total cross section is $0.77\pb$ at $\sqrtHatS=680\gev$, and it has similar distributions with the NS scenario. The total cross section is just $\sim14\%$ lower than the NS scenario. The production of $\tilde{\chi}^+_1\tilde{\chi}^-_2$ and $\tilde{\chi}^+_2\tilde{\chi}^-_2$ pairs requires a higher c.m. energy, and they are not reachable at a collider with $\sqrtHatS=1\tev$.

%\textbf{comparison between NS and mSUGRA}
     Comparing the NS and the mSUGRA scenarios shows that the total distributions of the cross sections for the same chargino pair are very similar to each other, the virtual corrections also have a similar trend as a function of the c.m. energy. Therefore, the cross section could not be enough to distinguish these two scenarios from each other. The $\hat{\sigma}^\text{LO+NLO}_\text{RR}$ and $\hat{\sigma}^\text{LO+NLO}_\text{RL}$ have the same trends as a function of c.m. energy on each of the scenarios given in Figs. \ref{fig9} (left) and (right).
      
      In the NS and the mSUGRA scenarios, the $\tilde\chi_1^+ $ decays primarily through the same channels as the RNS scenario.
      The sum of these branching ratios is again $\sim 0.999$. The decay channels are     
        $BR(\tilde\chi_1^+ \rightarrow   \tilde\chi_1^0 + u + \bar{d}) = 0.333 $, 
        $BR(\tilde\chi_1^+ \rightarrow   \tilde\chi_1^0 + c + \bar{s}) = 0.333 $, 
        and $BR(\tilde\chi_1^+ \rightarrow   \tilde\chi_1^0 + l + \nu_l ) = 0.333$. 
     The pattern at the detector is the same with the RNS scenario; there will be a missing-energy plus two light quark jets or a lepton at the final state for each of the charginos.

%    \textbf{MSUGRA, the second chargino}
%      2.16464967E-01      1000022        24        # W2SS+  -->  Z1SS   W+                 
%      2.50894636E-01      1000023        24        # W2SS+  -->  Z2SS   W+                 
%      2.45451689E-01      1000024        23        # W2SS+  -->  W1SS+  Z0                 
%      2.57460952E-01      1000024        25        # W2SS+  -->  W1SS+  HL0                

%        \begin{widetext}
%        \onecolumngrid
 %       \end{widetext}

\subsection{Brummer-Buchmuller (BB) benchmark scenario} 

    Inspired by GUT-scale string compactifications, Brummer and Buchmuller proposed this scenario in Ref. \cite{Brummer:2011yd, Brummer:2012zc} where the Fermi scale emerges as a focus point. In this scenario, gauge-mediated soft terms are characterized by discrete numbers, and they are called the messenger indices. 
    For certain models, the messenger indices are aligned such that the contributions to the mass of the Z boson cancel between various soft terms. Besides, the Higgs-doublet mixing parameter $\mu$ arises from the gravitational interactions. It was predicted that the graviton mass and $\mu$ are at the same order ($\mu\simeq m_{3/2}\simeq 150-200\gev$). In this study, the messenger indices are $(N_1, N_2, N_3)=(46,46,20)$, the gauge-mediated soft mass per messenger pair is $m_{\mathrm{GM}}=250\gev$, the ratio of the Higgs vacuum expectation value is taken as $\tan\beta = 48$, the Higgs-doublet mixing parameter is $\mu=167\gev$, and $m_A=4050\gev$ is set. More detailed information was presented in Refs. \cite{Brummer:2012zc, Baer:2013ula}. The results for the lightest chargino pair production in a $e^+e^-$-collider with detector simulation at the ILC was presented in Ref. \cite{Berggren2013}. Accordingly, this benchmark scenario was specifically adapted for the future lepton collider studies.
    
\begin{figure}[htbp]
        \centering
        \includegraphics[width=0.48\textwidth]{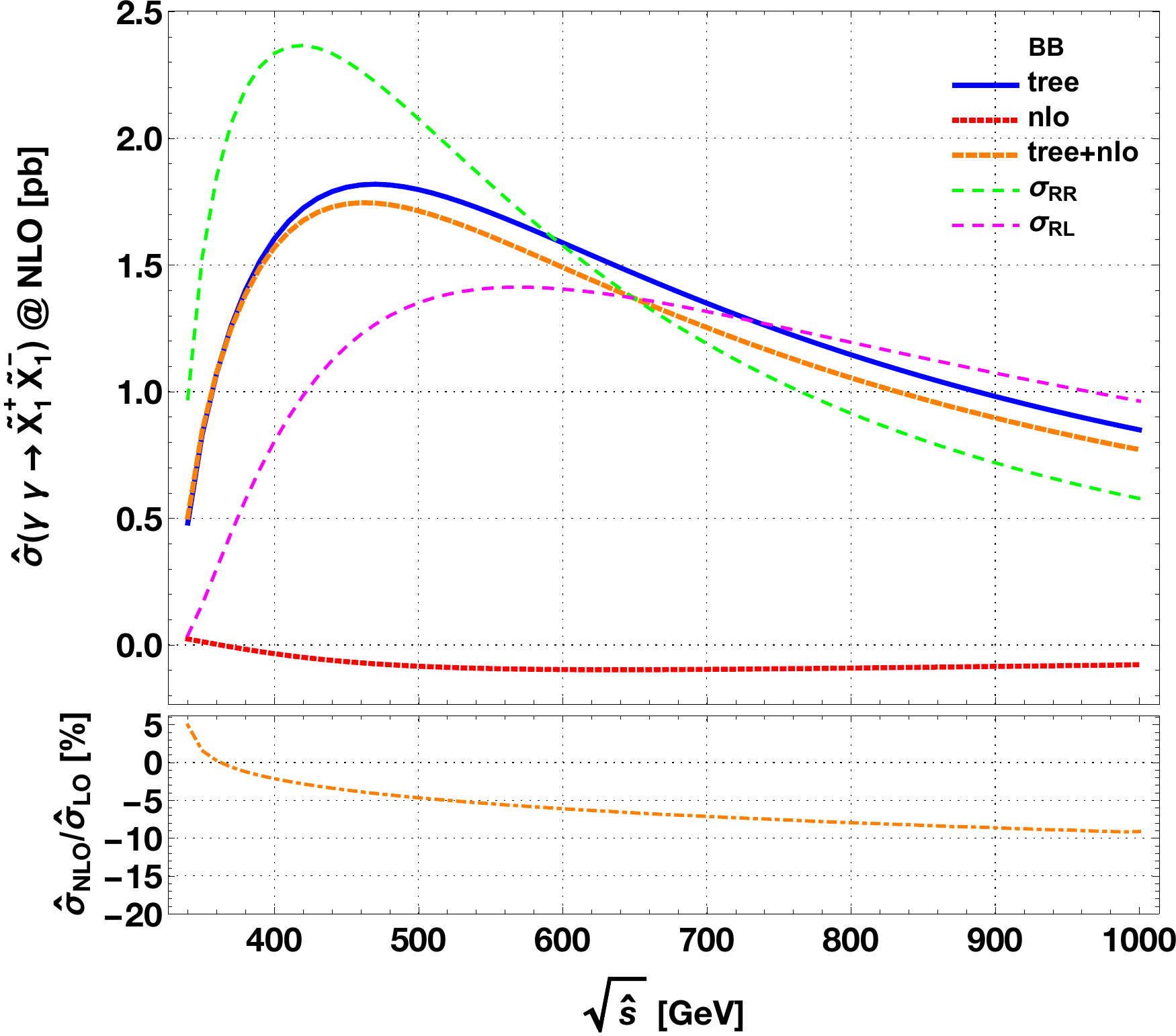}     
        \includegraphics[width=0.48\textwidth]{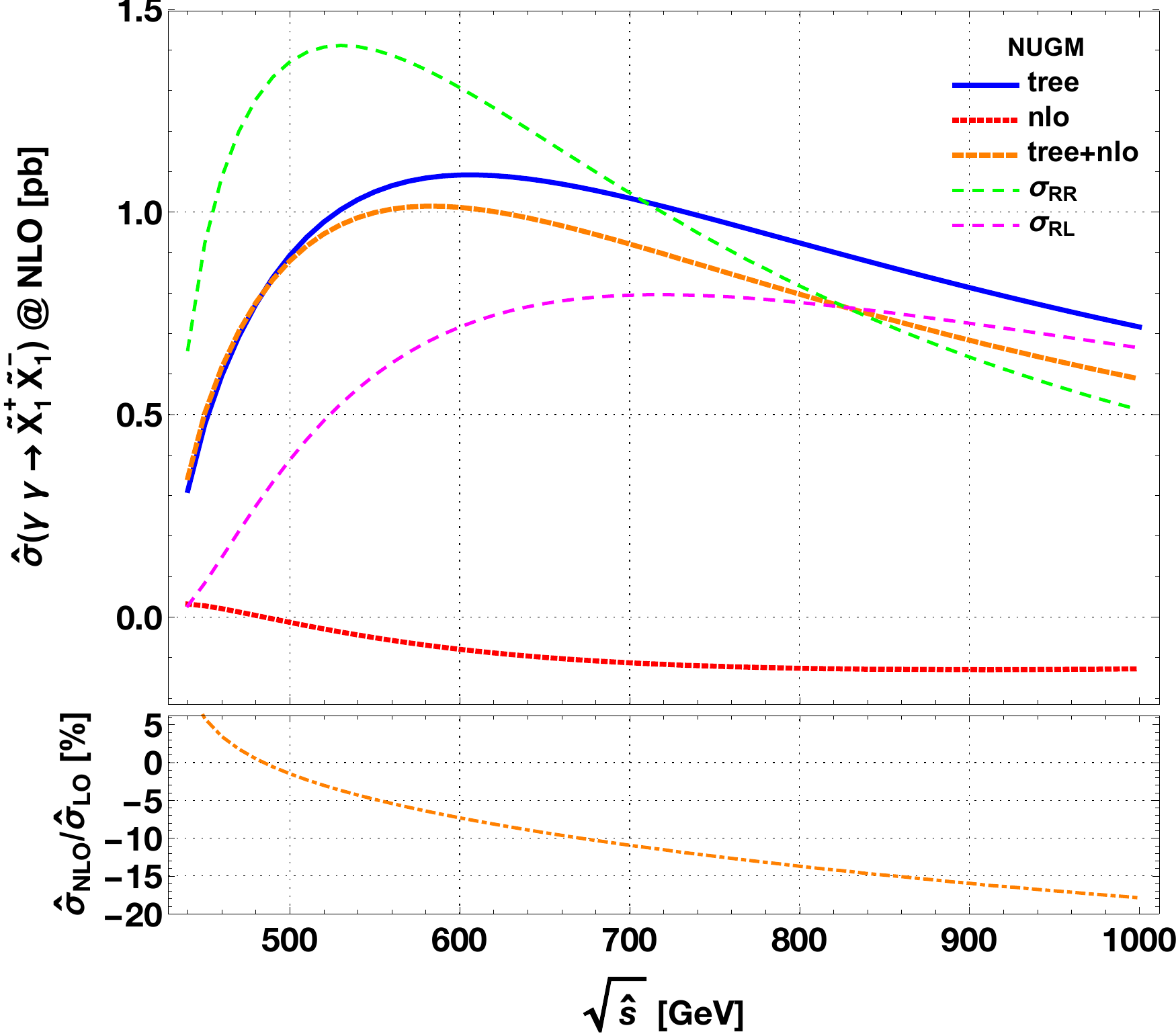}
        \caption{The cross section of the process $\gamma\gamma\rightarrow\tilde\chi_i^+\tilde\chi_j^-$ and the ratio $\hat{\sigma}_\text{NLO}/\hat{\sigma}_\text{LO}$ as a function of $\hat{s}$ are plotted. Two polarization cases $\hat{\sigma}^\text{LO+NLO}_\text{RR}$ and $\hat{\sigma}^\text{LO+NLO}_\text{RL}$ are indicated by green-dotted and magenta-dotted lines in the figure, respectively. (left): BB scenario, and (right): NUGM scenario.}
        \label{fig10}
        \end{figure}

    The lightest chargino mass is around $m_{\tilde{\chi}_{1}^{\pm}}\sim 167\gev$, but the second lightest chargino mass is at the order of ten TeV. Therefore, only the lightest chargino pair is accessible in a $\gamma\gamma$-collider with $\sqrt{s}\leq1\tev$. The energy dependence of the total cross section of $\hat{\sigma}(\gamma\gamma\rightarrow\tilde{\chi}^+_1\tilde{\chi}^-_1)$ is given in Fig. \ref{fig10} (left). The total cross section reaches up to $1.75\pb$ at $\sqrt{\hat s}=460\gev$, and that is the second highest cross section after the RNS scenario. 
    The ratio $\hat{\sigma}_\text{NLO}/\hat{\sigma}_{LO}$ gets down to $-9.13\%$ at $\sqrtHatS=1\tev$. 
    Also, the polarized cross section of $\hat{\sigma}^\text{LO+NLO}_\text{RR}$ rises up to $2.37\pb$ given in Fig. \ref{fig10} (left).

    Due to the mass split between the lightest chargino and the lightest neutralino ($m_{\tilde{\chi}^+_1}-m_{\tilde{\chi}^0_1}$) is at the order of $\approx 770\text{ MeV}$, the $m_{\tilde{\chi}^+_1}$ mainly decays through charged pion and neutralino. Besides, the decay channels of the lightest charginos in a collider at the ILD were studied in \cite{Berggren2013, Sert:293539}. The branching ratios are $BR(\tilde\chi_1^+ \rightarrow \tilde\chi_1^0 + e^+ +\nu_e ) = 0.15$, $BR(\tilde\chi_1^+ \rightarrow   \tilde\chi_1^0 + \mu^+ + \nu_\mu ) = 0.137$, and $BR(\tilde\chi_1^+ \rightarrow   \tilde\chi_1^0 + \pi^+ ) = 0.604$. 
    %At last, the possible polarization cases of the photon beams coming to the collision are examined. The same distribution given in Fig. \ref{fig9} is obtained, and the ratio $\sigma(P_{\gamma},P_{\gamma})/\sigma_{UU}$ reaches up to 1.2 at the top-left and bottom-right corners in BB scenario. 

\subsection{Non-universal gaugino masses (NUGM) scenario}

    %Finally, the computation for the lightest chargino pair is also carried out for the NUGM scenario.
    %\textbf{NUHM2 Scenario}\\
    %The two-parameter non-universal Higgs model (NUHM2) was inspired by GUT models where $\hat{H}_u$ and $\hat{H}_d$ belong to different multiplets. Accordingly, the soft-breaking scalar masses of the two Higgs doublets $m^2_{{H}_u}$ and $m^2_{{H}_d}$ are taken as a free parameter. Since there is no driving force to assume that the Higgs fields and sfermion fields unify at the GUT scale, the sfermions and the two Higgs doublet fields are assumed that they have a different universal soft-breaking scalar mass terms. It is argued that in Ref. \cite{Baer_2005}, the GUT scale masses $m^2_{{H}_u}$ and $m^2_{{H}_d}$ could be trade-off for the weak scale parameters $\mu$ and $m_A$. 
    %The \textsc{SLHA} files are calculated using \textsc{ISASUGRA}, and the GUT scale parameters are set as follows: $\mu=6\tev$ and $m_A=275\gev$, and the rest of the parameters are given in Tab. \ref{tab:rnsparam}. 
    %As a result, the masses of all the Higgses and the elektroweakinos become light at the order of $100-300\gev$, but the rest of the sparticles are still beyond the TeV, so does beyond the reach of the LHC. The distribution of the total cross section is given in Fig. \ref{fig11} (center), and it reaches to a maximum of $0.83\pb$ at $\sqrtHatS=0.64\tev$. The sum of the virtual and the real corrections shares the same trend as the other scenarios, and it goes down to $-16.4\%$.
 
    %\textbf{NUGM Scenario}\\
    The non-universal gaugino masses (NUGM) scenario is motivated by the GUT models where the universality of the gaugino masses are loosened at $M_{GUT}$. Besides, that also settles the little hierarchy problem \cite{Baer:2013ula}. This model is beyond the reach of the LHC because the mass spectrum is at the order of TeV. However, the masses of the charginos and the lightest two neutralinos are set to be around $100-250\gev$. The mass spectrum and the relevant parameters at the weak scale are calculated with \textsc{ISASUGRA} using the input parameters given in Tab. \ref{tab:rnsparam}. 
    The total production cross section is $\sigma_{\tilde{\chi}^+_1\tilde{\chi}^-_1}=1.01\pb$ at $\sqrtHatS=0.58\tev$. 
    At $\sqrtHatS>0.49\tev$, the sum of the virtual and the real corrections becomes negative, and the tree-level cross section is lowered by $16.2\%$. 
    In this scenario, the lightest chargino decays mainly via the lightest neutralino and a W-boson with $BR(\tilde\chi_1^\pm \rightarrow  \tilde\chi_1^0  + W^\pm) = 0.999$. W-boson decaying to light quarks or lepton + neutrino will give the same final states with the previous scenarios, but the W-boson is not virtual in this scenario.
    
    Reconstructing large missing energy plus two W-bosons at the detector is enough to extract the chargino pair. However, the scattering of $\gamma\gamma\rightarrow W^+W^-$ has a substantial cross section that will be discussed later. Still, this channel could be promising for distinguishing the lightest chargino pair from the background in a $\gamma\gamma$-collider. The production of $\tilde{\chi}^+_1\tilde{\chi}^-_2$ and $\tilde{\chi}^+_2\tilde{\chi}^-_2$ are kinematically not possible with $\sqrt{s}=1\tev$ for the NUGM scenario. A similar study with the same final state was carried out in Ref. \cite{Klamke:2005jj} employing full detector simulation.
%        \begin{widetext}
%        \onecolumngrid
%        \end{widetext}

    %Therefore, the possible background channels need to be taken into account considering the acceptance and the instrumentation of the detector to access the NUGM scenario fully.       

    \subsection{Polarization of the incoming beams}
    
\begin{figure}[htbp]
        \centering
        \includegraphics[width=0.480\textwidth]{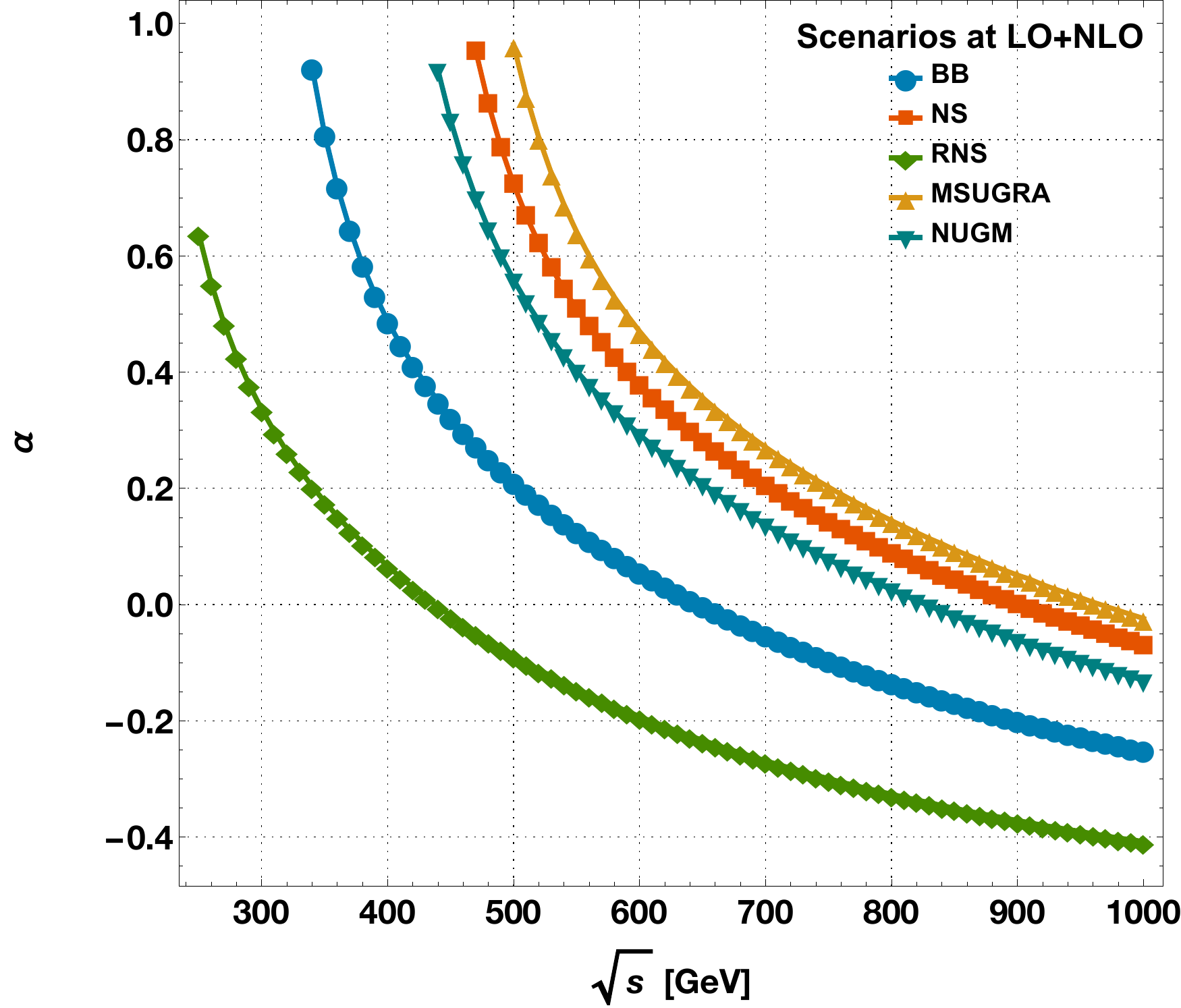}
        %\includegraphics[width=0.450\textwidth]{figures/plot11_ratio_alpha_pp2}
        %\hfill
        %\includegraphics[width=0.30\textwidth]{plot12}
        \caption{\label{fig11} The distribution of the parameter $\alpha$ as a function of the c.m. energy for various scenarios.}
        \end{figure}

    The polarization of the incoming particles could enhance or suppress the contribution coming from various Feynman diagrams given in Fig. \ref{fig1} - \ref{fig6}. The cross section for various polarization cases could be calculated with the relation given below:
    \begin{equation}
    \label{eq12}
    \hat{\sigma}(P_{\gamma^1},P_{\gamma^2})=\frac{1}{4} \sum_{\alpha,\beta=\pm1} (1+\alpha P_{\gamma^1})(1+\beta P_{\gamma^2})\hat{\sigma}_{\alpha\beta} ,
    \end{equation}
    where $\hat{\sigma}_{LR}$ stands for the cross section with the photon beam is polarized completely left-handed ($P_{\gamma} = -1=L$), and the photon beam is polarized completely ($P_{\gamma} = +1=R$) right-handed. Accordingly, the cross sections $\hat{\sigma}_{RL}$, $\hat{\sigma}_{LL}$, and $\hat{\sigma}_{RR}$ are defined similarly. 
    In a $\gamma\gamma$-collider, the following relations hold between the cross sections: $\hat{\sigma}_{LL}=\hat{\sigma}_{RR}$ ($J_z=0$), and $\hat{\sigma}_{RL}=\hat{\sigma}_{LR}$ ($J_z=2$). Then, the ratio $\hat{\sigma}(P_{\gamma},P_{\gamma})/\hat{\sigma}_{UU}$ becomes   
        \begin{equation}
        \label{eq13}
        \frac{\hat{\sigma}(P_{\gamma^1},P_{\gamma^2})}{\hat{\sigma}_{UU}}= 1+\alpha P_{\gamma^1} P_{\gamma^2}   ,
        \end{equation}
    where $\alpha=0.5(\hat{\sigma}_{RR}-\hat{\sigma}_{RL})/\hat{\sigma}_{UU}$, and it is the only relevant parameter defines the ratio given in Eq. \ref{eq13}. The $\hat{\sigma}_{UU}$ represents the cross section calculated with the unpolarized incoming photon beams. 
    In Fig. \ref{fig11}, the parameter $\alpha$ is plotted including tree and loop level contributions as a function of the c.m. energy. The cross section as a function of the possible polarization configurations of the incoming photon beams ($P_{\gamma}$) could be calculated easily. In Fig. \ref{fig11}, the distributions show that for $\alpha>0$ the same helicity beam enhances the cross section while in $\alpha<0$ the opposite helicity raises the cross section.
    Depending on the c.m. energy, the cross section could be raised by 90\% for some of the scenarios. For example at $\sqrtHatS=0.55\tev$, the same polarizations of the photons increases the total cross section by $\sim40(\sim65)\%$ in the NUGM (MSUGRA) scenario. However, the the opposite polarization enhances the total cross section by $\sim40\%$ in the RNS at $\sqrtHatS=1\tev$.          
    Since NLO contribution is not dominant in the production of the lightest chargino pairs in each scenario, the parameter $\alpha$ is mainly ruled by the tree-level cross section. 

\subsection{Convoluted cross section in a $e^+e^-$-collider}
    
\begin{figure}[htbp]
    \centering
    \includegraphics[width=0.48\textwidth]{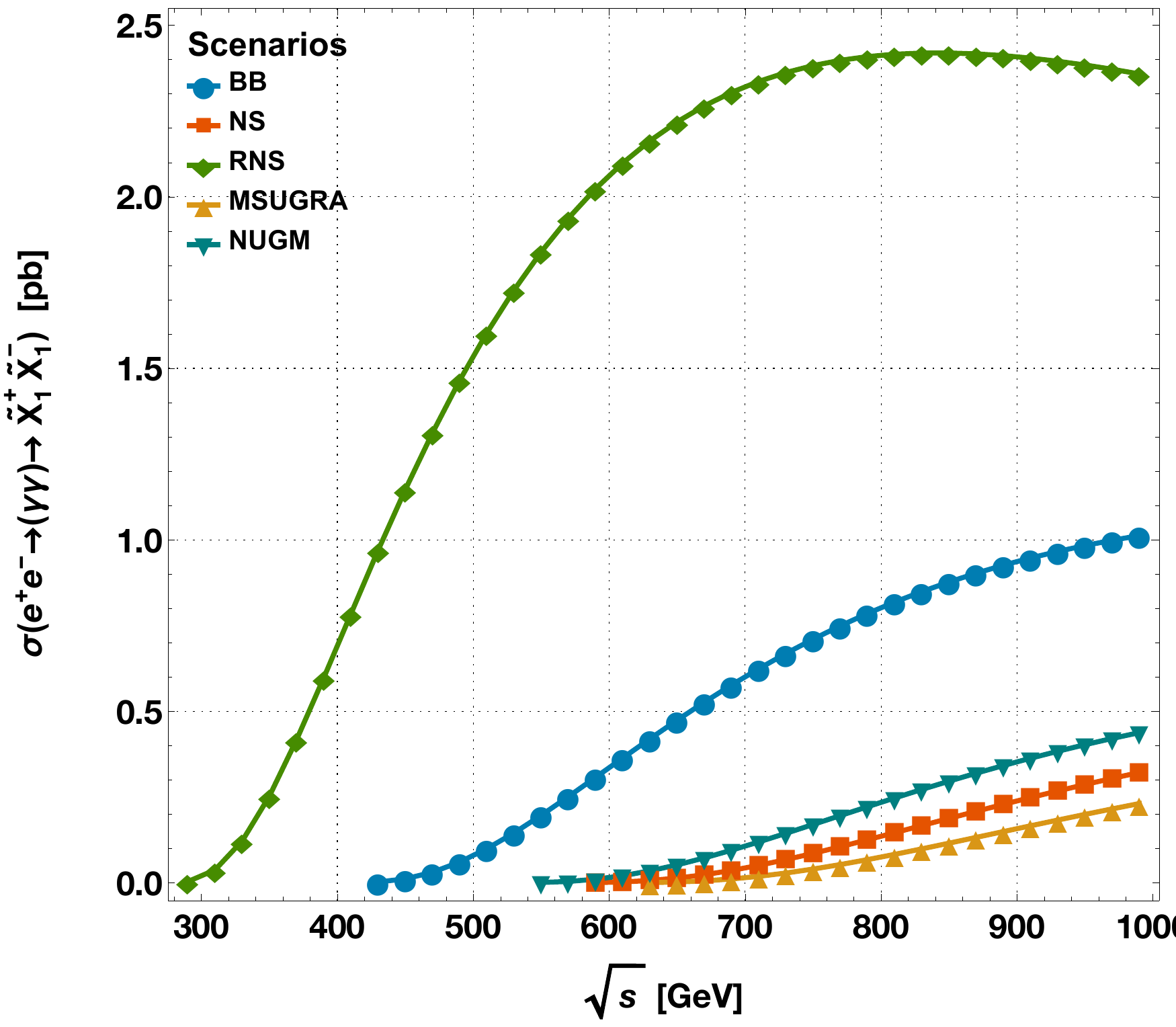}
    \includegraphics[width=0.48\textwidth]{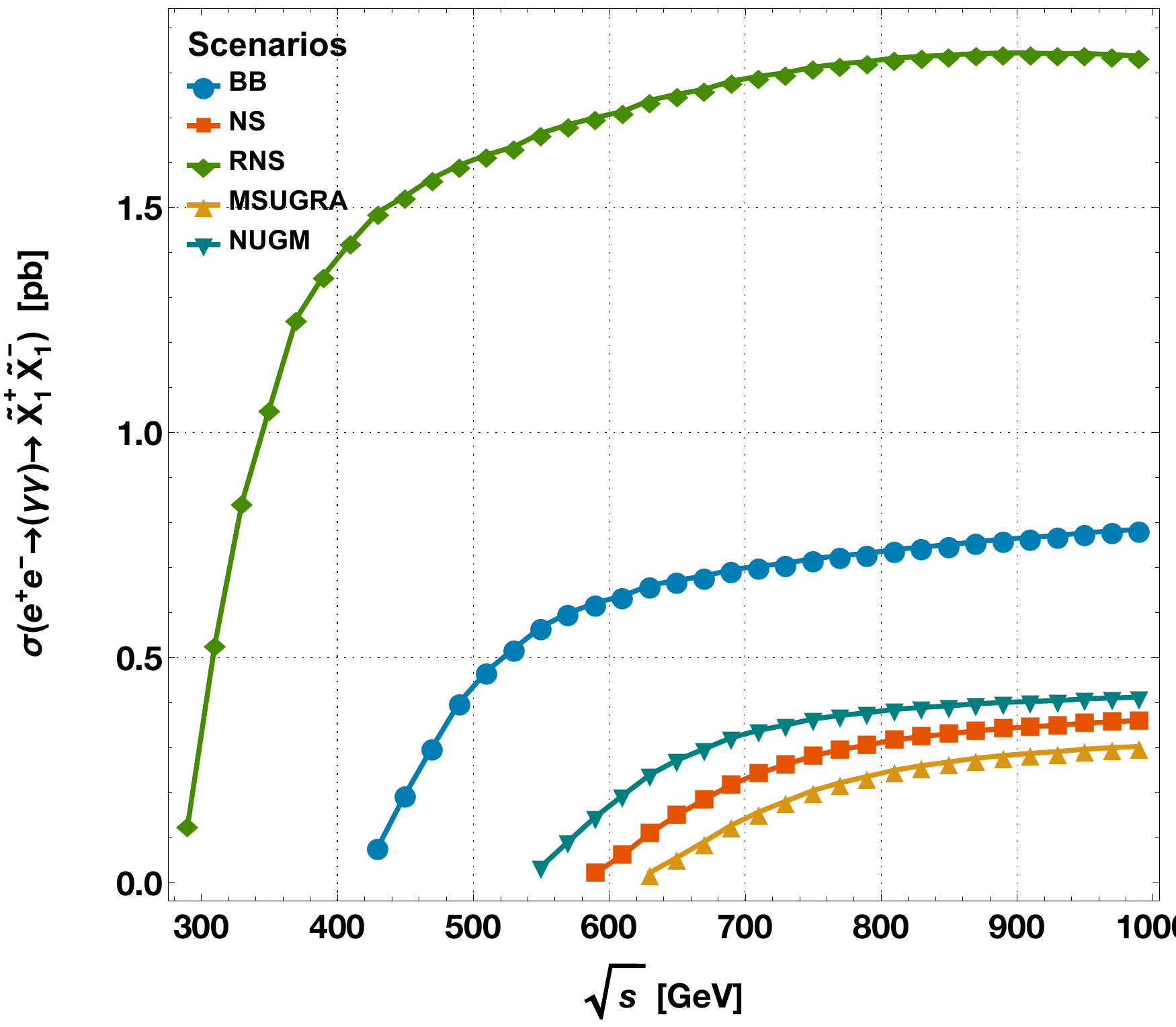}
    \caption{\label{fig12} The total cross section of the lightest chargino pair production convoluted with the photon luminosity in an $e^+e^-$-collider is given for the scenarios depicted in the figure. (left): $P_\text{el}=+0.8$ and $P_\text{l}=+1$, (right): $P_\text{el}=+0.8$ and $P_\text{l}=-1$ are assumed.}
    \end{figure}

    The total cross section convoluted by the photon luminosity in a $e^+e^-$-collider $(\sigma(e^+e^-\rightarrow\gamma\gamma\rightarrow \tilde{\chi}^+_1\tilde{\chi}^-_1))$ is presented in Fig. \ref{fig12} for each of the benchmark points. In photon collider, the possibility to obtain high degree of photon polarization is one of the main advantages. Therefore, we assumed the polarization of the laser beam to be $P_\text{l}=\pm1$, and the electron beam polarization as $P_\text{el}=\pm0.80$ for producing the Compton back-scattered photons. However, only the two interesting configurations are given in Fig. \ref{fig12}. 
    It is evident in the figure that the NUGM, the MSUGRA, and the NS scenarios have similar trends due to the similar partonic distributions. 
    The minimal c.m. energy required for a detection at a $e^+e^-$-collider is $\sqrt{s}\gtrsim 0.65 \tev$ for both $P_\text{el}$ and $P_\text{l}$ configurations. 
    The cross section rises with the c.m. energy as expected, and they get in the range of $238-446\fb$ at $\sqrt{s}= 1\tev$ in $P_\text{el}=0.8$ and $P_l=-1$. In fact, the cross sections do not change much in high energies for the other laser and electron polarization configurations because the peak in the energy spread of the photons gets narrow at higher c.m. energies, and also the two contributions $\hat{\sigma}_\text{RR}$ and $\hat{\sigma}_\text{RL}$ are very close to each other at $\sqrt{s}= 1\tev$ in the NS, the MSUGRA, and the NUGM scenarios.
    The rise of the convoluted cross section in these scenarios is expected due to the increase towards $\sqrtHatS\approx0.5\tev$ in the partonic cross section of these scenarios. Therefore, the convoluted cross section rises in Fig. \ref{fig12} for these scenarios. The second highest convoluted cross section is obtained for the BB scenario, and it reaches up to $1\pb$. 
    The highest convoluted cross section is obtained for the RNS scenario as expected because the partonic cross section is the highest one. 
    In the RNS scenario, the process becomes accessible  even as low as $\sqrt{s} \sim 0.30\tev$ with $\sigma_{\tilde\chi_1^+\tilde\chi_1^-}=0.309\pb$, and it rises quickly above $1.70
    \pb$ for $\sqrt{s}\geq0.6\tev$ with $P_\text{l}=-1$ and $P_\text{el}=+0.8$.
    
    %At higher c.m. energies, due to the narrowed energy range (but still high) of the photons and the $1/s$ suppression of the partonic cross section, it reaches to saturation, then falls slowly. 
    
    Next, comparing the two polarization configurations given in Fig. \ref{fig12} reveals one of the advantages of the $\gamma\gamma$-collider that could be seen at low energies. The number of high energy photons is increased significantly when the helicity of the electron and the laser are opposite compared to the other configurations.

    Therefore, the convoluted cross section is boosted more than twice between $P_\text{l}=+1$ and $P_\text{l}=-1$ configurations (Fig. \ref{fig12} (right)) in the RNS scenario at $\sqrtHatS=0.4\tev$, and a similar trend could be seen in the other scenarios. After a sharp rise, the convoluted cross section displays saturation, and it varies slowly. Contrary to that the convoluted cross section gets higher values for all the scenarios plotted in Fig. \ref{fig12} (left) with $P_\text{l}=+1$ configuration. 

%    \begin{widetext}
%    \onecolumngrid
%    \end{widetext}

\subsection{Possible background channels}

    The production rate of the charginos has a moderate cross section that has a range of $0.77-3.94\pb$ depending on the scenario.  The possible background channels in a $\gamma\gamma$-collider need to be discussed to assess the observability of the signal. The decay chains of the lightest chargino for each of the scenarios are presented with the results. According to them, the identification of the chargino requires reconstruction of large missing energy due to the LSP which appears in chargino decay in all of the scenarios.  Then, depending on the decay chain, two light quarks or a lepton plus a neutrino (which is another source for the missing energy) will be produced for each chargino in the RNS, the NS, the mSUGRA, and the BB scenarios. As a result, the signal at the detector will be as follows: large missing energy + two leptons, a large missing energy + 4 jets, and a large missing energy + 2 jets + a lepton. 
    In the NUGM scenario, the mass split ($m_{\tilde{\chi}^+_1}-m_{\tilde{\chi}^0_1}$) is larger than the mass of the W-boson, and therefore the produced W-boson is not virtual like in the other scenarios, which could be used to eliminate the fake events in the detector.  
        
    According to the Ref. \cite{Brinkmann:1997sj}, the following scattering processes in a $\gamma\gamma$-collider have a pretty large cross section, and they are potentially the primary background channels:  
            $\gamma\gamma \rightarrow  \mu^+ \mu^- \mu^+ \mu^- $ and 
            $\gamma\gamma \rightarrow  e^+ e^- (\mu^+\mu^-)$ \cite{Miller:1994xw, Richard:1994vn}. However, requiring only two leptons and large missing energy at the final state could cancel the large portion of these two channels. 
    Besides, the scattering processes $\gamma\gamma \rightarrow  W^+ W^- $ \cite{Yasui:1992fh} and $\gamma\gamma \rightarrow  W^+ W^- \gamma $ have also large cross sections. These channels could be eliminated by employing a cut on the transverse missing energy and requesting no photon at the final state. In any case, the miss-reconstruction of the events could easily contaminate the events considered as the signal. 

    One other important background process is $\gamma\gamma \rightarrow  W^+ W^- Z^0$ \cite{Ginzburg:1993gz} with the cross section about 1 $\pb$. If Z-boson decays through $Z\rightarrow \nu\bar{\nu}$, then these neutrino pairs with the W-boson pair at the final state could imitate the signal. This process could be considered as one important background. 
           
     Another set of possible background processes is the production of hadrons at the final state. One of the main final states consists of large missing energy + four jets. The cross section of $\gamma\gamma \rightarrow  {\text{hadrons}}$ is $\approx 0.4-0.6 \ub$ \cite{Chen:1993dba, Brinkmann:1997sj}, and the processes with diffraction physics in the soft region have $\sigma_{tot}\sim 0.5-1 \ub$ at $W_{\gamma\gamma} \sim 2 \tev$ \cite{Chen:1993dba}. Even for the hadronic final state the most serious background will be $\gamma\gamma \rightarrow  W^+ W^-$. However, the W-boson pair coming from the background will have a back-to-back topology, therefore, they could be suppressed by asking the W-bosons to be scattered by a large acoplanarity. Besides, pure QCD events have a flat distribution, that could also be useful to eliminate the hadronic background events. On the other hand, since there are 4 jets for one of the final states, jet finding algorithms will be significant to extract the chargino pair production. 
     
    Overall, it could be much easier and cleaner to extract the chargino pair signals from the events having a large missing energy + 2 leptons compared to large missing energy + four light jets. A better assessment of the pattern of $\tilde{\chi}_1^\pm$-pairs in the detector most certainly requires Monte Carlo simulation including all the potential background processes. To this end, the instrumentation, the acceptance and the efficiency of a detector must be included. That is beyond the scope of this study.

    %\textbf{The branching ratio of the charginos:}
    %\rred{In a detector, the lightest chargino follows three-body decay for the RNS, NS, mSUGRA scenarios, and the topology of the signal at the detector is large missing energy from the dark matter candidate (the lightest neutralino) + two jets or a lepton for each chargino.} Moreover, the decay of the lightest chargino in the NUHM2 and the NUGM scenarios differs from the previous scenarios, and the chargino mainly decays via the lightest neutralino and a W-boson with $BR(\tilde\chi_1^+ \rightarrow  \tilde\chi_1^0 + W^+ = 0.999)$. Then, the signal at the detector would be large missing energy + two W-bosons. In the BB scenario, the $\tilde\chi_1^\pm$ are mainly decayed through $\tilde\chi_1^0$ and a pion. 

%%%%%%%%%%%%%%%%%%%%%%%%%%%%%%%%%%%%%%%%%%%%%%%%%%%%%%%%%%%%%%%%%%%%%%%%%
%%%%%%%%%%%%%%%%%%%%%%%%%%%%%%%%%%%%%%%%%%%%%%%%%%%%%%%%%%%%%%%%%%%%%%%%%

\section{Conclusion and Summary}
\label{sec5}

    In this study, the lightest chargino pair production in $\gamma\gamma$ collisions is investigated. 
    The results include all the virtual, the soft, and the hard QED corrections for various benchmark points and scenarios.
    %\textbf{compare the benchmark points:}
    The presented numerics show that the RNS has the highest total cross section between all the scenarios considered in this paper, and it reaches up to $3.9\pb$. 
    Besides the RNS, the BB scenario has an elevated cross section. 
    The other scenarios such as the NS, the mSUGRA, and the NUGM have identical distributions at the tree-level and small variations at the NLO-level. 
    The sum of the virtual and the real photon corrections is also similar, and they reach a maximum of $-10\%$ at $\sqrt{\hat{s}}=1\tev$.      The contributions coming from $\hat{\sigma}^\text{LO+NLO}_\text{RR}$ and the $\hat{\sigma}^\text{LO+NLO}_\text{RL}$ to the total cross section are plotted for each of the scenarios. It is presented that the $J_z=0$ dominates at low energies and gets lower at increasing energies while the $J_z=2$ rises at the same time. Then, the $J_z=2$ becomes higher than the $J_z=0$ at higher energies. 

    The NUGM has the highest $\hat{\sigma}_{NLO}/\hat{\sigma}_{LO}$ ratio meaning that the one-loop corrections become important for this scenario. 
    The total cross section as a function of the incoming photon polarization is analyzed for each of the scenarios. The distributions of the parameter $\alpha$ show the contribution of the same and the opposite photon polarization configurations to the total cross section for a given c.m. energy. 
     
    %All the possible polarization cases of the incoming beams are scanned, and the change of the cross section is presented to see the potential of such a collider if such an option becomes achievable.
    %The distributions are identical as expected in all the scenarios for the given c.m. energy because the process is a QED at the tree level.
    %The cross section is enhanced up to for left-handed polarized photon beam and right-handed polarized photon beam at $\sqrtHatS=1\tev$. 
    %The enhancement is around 30\% in the NUGM and 20\% in the BB scenario. The polarization of the photons does not have a significant impact on the mSUGRA, and the NS scenarios at.
    
   %\bblue{Considering the highest ones, they reach up to $2.4\pb$, $1.2 \pb$, and $0.4\pb$ at $\sqrt{s}=0.5\tev$ for, respectively.}   
   %\bblue{The highest ones reach up to $2.4\pb$, $1.2 \pb$, and $0.4\pb$ at $\sqrt{s}=0.5\tev$ for the RNS, the NUGM, and the BB scenarios, respectively.} 

    The total cross section in a $e^+e^-$-collider is calculated by convoluting the process $\gamma\gamma \rightarrow \tilde{\chi}^+_i\tilde{\chi}^-_j$ with the photon luminosity, and the RNS scenario has the highest value among all the scenarios. The RNS and the BB scenarios have a convoluted cross section of $1.61 \pb$, and $0.43\pb$ at $\sqrt{s}=0.5\tev$, respectively, and these two scenarios could be accessed at the ILC. Moreover, it is possible to test the lightest chargino pair production in the RNS scenario with as less energy as $\sqrt{s}=350\gev$. However, the NUGM, the NS, and the MSUGRA scenarios require higher c.m. energy ($\sqrt{s}>0.6\tev$).

    %\textbf{The ending:}
    If the LHC could not find any hint for the supersymmetry, a significant portion of the parameter space would be excluded. However, there will be some region left where the supersymmetry is not tested yet. The future $e^+e^-,\;\mu^+\mu^-$ or $\gamma\gamma$ colliders could rule out those parameter spaces and put the supersymmetry at rest, or discover the supersymmetric particles. The results manifest the potential of the $\tilde\chi_1^+\tilde\chi_1^-$ production in a $\gamma\gamma$-collider, and that could be used for possible optimization in accelerator and detector design. The FLC with $\gamma\gamma$-collider mode is an ideal laboratory to study the supersymmetry. It will complement the results of the LHC with high precision, and it could also lead to discoveries that reveal the secrets of the universe.
     
%%%%%%%%%%%%%%%%%%%%%%%%%%%%%%%%%%%%%%%%%%%%%%%%%%%%%%%%%%%%%%%%%%%%%%%%%
\section*{Acknowledgements}
The computation presented in this paper is partially performed at TUBITAK ULAKBIM, High Performance and Grid Computing Center (\textsc{TRUBA} resources), and also at the computing resource of \textsc{FENCLUSTER} (Faculty of Science, Ege University). Ege University supports this work, project number 17-FEN-054. This work would not have been possible without the time I borrowed from Cinar Sonmez.

%%%%%%%%%%%%%%%%%%%%%%%%%%%%%%%%%%%%%%%%%%%%%%%%%%%%%%%%%%%%%%%%%%%%%%%%%
\bibliography{template-8s_revtex}

\end{document}